\documentclass{article}

\usepackage{type1cm}        
\usepackage{makeidx}         
\usepackage{graphicx}        
\usepackage{tabularx}                         
\usepackage{multicol}        
\usepackage[bottom]{footmisc}

\usepackage{newtxtext}       %
\usepackage{newtxmath}       

\usepackage[nolist]{acronym} 

\usepackage[round]{natbib}

\usepackage{authblk}

\graphicspath{{figures/}}

\makeindex             

\providecommand{\keywords}[1]
{
  \small	
  \textbf{\textit{Keywords---}} #1
}

\begin{document}

\title{IEC-61850 Performance Evaluation in a 5G Cellular Network: UDP and TCP Analysis}

\author[1]{Iurii Demidov}
\author[1]{Dick Carrillo Melgarejo}
\author[1]{Antti Pinomaa}
\author[2]{Liana Ault }
\author[2]{Jari Jolkkonen}
\author[2]{Kirsi Leppa}
\affil[1]{School of Energy Systems, LUT University, Yliopistonkatu 34, 53850 Lappeenranta, Finland}
\affil[2]{NOKIA FI, Finland}

%
%
\maketitle

\abstract{
This chapter summarizes the results obtained from a test bed, which is composed of a microgrid and a wireless network, both relying on real and practical premises. 
This test bed aims to evaluate application protocols considered in the transport layer of the standard IEC-61850. 
The application is a data gathering system, where IEC-61850 messages are transmitted from and between different elements in a microgrid system, such as \acsp{IED}, publishers, and subscribers, which are node elements that are part of the utility grid.
The standard IEC-61850 defines many protocols, such as \acs{SV}, \acs{GOOSE}, and \acs{MMS}, each of them with a variety of requirements and sharing similar transport communication protocols, such as \acs{TCP} and \acs{UDP}. 
Hence, a testing framework is used in the test bed in order to evaluate the performance of these protocols aiming to enable the communication in a real microgrid deployment.
The IEC-61850 messages transported by \acs{TCP} or \acs{UDP} are transmitted  to a centralized location, where a local database stores the information under different test scenarios and using different transmission sample times.
The overall results are promising and indicate that the protocols based on \acs{TCP} in combination with a \acs{5G} cellular network architecture would be suitable for most applications within the studied microgrid case. In the case of \acs{UDP}, the performance indicated some specific constraints related to the multiple access nature of simultaneous transmissions and also based on the \acs{SINR} .
}

\begin{acronym}
  \acro{1G}{first generation of mobile network}
  \acro{1PPS}{1 pulse per second}
  \acro{2G}{second generation of mobile network}
  \acro{3G}{third generation of mobile network}
  \acro{4G}{4th generation of mobile network}
  \acro{5G}{5th generation}
  \acro{ARQ}{automatic repeat request}
  \acro{ASIP}{application specific integrated processors}
  \acro{AWGN}{additive white Gaussian noise}
   \acro{BER}{bit error rate}
   \acro{IP}{Internet protocol}
   \acro{MAC}{media access control}
   \acro{LTE}{long-term evolution}
   \acro{EPS}{Evolved Packet System}
   \acro{EPC}{Evolved Packet System}
   \acro{E-UTRAN}{Evolved UTRAN}
   \acro{eNB}{evolved Node B}
   \acro{OFDMA}{Orthogonal Frequency Division Multiple Access}
   \acro{SC-FDMA}{Single Carrier Frequency Division Multiple Access}
   \acro{3GPP}{3rd Group Partnership Project}
   \acro{SNR}{signal to noise-ratio}
   \acro{BPSK}{binary phase shift keying}
   \acro{AWGN}{additive white Gaussian noise}
   \acro{PER}{packet error ratio}
   \acro{WLLN}{weak law of large numbers}
   \acro{pdf}{probability density function}
   \acro{MCS}{modulation and coding scheme}
   \acro{UE}{user equipment}
   \acro{PGW}{packet data network gateway}
   \acro{SGW}{serving gateway}
   \acro{PHY}{physical layer}
   \acro{HARQ}{hybrid automatic repeat request }
   \acro{RLC}{Radio link control}
   \acro{PDCP}{Packet data convergence protocol}
   \acro{ROHC}{robust header compression}
   \acro{RRC}{radio resource control}
   \acro{SIB}{system information block}
   \acro{NAS}{non-access stratum}
   \acro{MME}{mobility management entity}
   \acro{CEI}{consumer-end inverter}
   \acro{IPC}{industrial personal computer}
   \acro{PV}{photo-voltaic}
   \acro{FO}{fibre optic}
   \acro{BESS}{battery energy storage system}
   \acro{RS}{rectifier substation}
   \acro{ICT}{information and communication technology}
   \acro{IT}{terrian-isolated functionalities unearth grounding}
   \acro{KPI}{key performance indicator}
   \acro{CPE}{customer premise equipment}
   \acro{IED}{industrial electronic device}
   \acro{PDN}{packet data network}
   \acro{MMS}{manufacturing message specification}
   \acro{GOOSE}{generic object-oriented substation event}
   \acro{SV}{sampled value}
   \acro{MIMO}{multiple input multiple output}
   \acro{EPC}{evolved packet core}
   \acro{ICMP}{Internet control message protocol}
   \acro{UDP}{user datagram protocol}
   \acro{TCP}{transmission control protocol}
   \acro{RSRP}{reference signal received power}
   \acro{RSRQ}{reference signal reference quality}
   \acro{SINR}{signal to noise ratio}
   \acro{RSSI}{received signal strength indication}
    \acro{WAN}{wide area network}
    \acro{RES}{renewable energy source}
    \acro{DSO}{distribution system operator}
    \acro{RAN}{radio access network}
    \acro{LAN}{local area network}
    \acro{LVDC}{low-voltage direct current}
\end{acronym}

\section{Introduction}

The impact of cellular networks on society has been defined by the evolution of wireless communication technologies, which have been applied for many generations.
Most of these developments have focused on human applications. However, nowadays, cellular networks aim to support many industrial or vertical applications, such as energy systems, grid automation, mining industry, and process automation industry.

\subsection{Evolution of Power System Industry}
Traditional power grids with unidirectional power flow are being transformed into bidirectional flow, which is related to the increasing amount of distributed generation and energy storage units connected to the grid. 
In other words, the traditional end consumers are being replaced by prosumers, which are end users that both consume and generate energy.
These new elements have brought up attractive and interesting use case scenarios in microgrid systems.
For instance, some parts of electricity distribution networks that are connected to grids are able to run in the islanded mode (a microgrid is said to be in the islanded mode when it is disconnected from the main grid and it operates independently with micro sources and load).
In other cases, \acp{RES} in power grids, such as weather-dependent wind and solar power generation, as well as the energy storage distributed to power systems, increase the need for having more monitoring, control, and protection applications that support the requirements of grid automation.
Accordingly, these grid applications set more stringent requirements for the communication networks enabling data exchange between different nodes in the grid and between the grid and the remote systems of \acp{DSO}.

Nowadays, these applications are run on wired networks, such as power line communication and fibre optic networks, which are mostly based on the Ethernet protocol. 
For instance, fibre optic networks are used in the automation of substations and also to connect two different substations, in which the communication network is used as a backhaul between networks, for example for connectivity between two remote monitoring rooms of two different \acp{DSO}.

\subsection{Evolution on Cellular Network Industry}
Likewise, the power grid evolution in the last decade, that is, the \ac{4G}, revolutionized the cellular network industry because it was the first system supporting an end-to-end packet-switched network; for instance, voice calls can be made using a full \ac{IP} network. 
The improvements of a variety of features in the protocol stack enabled connectivity between devices and applications on a diversity of industrial applications.
However, most of the critical applications are still being supported by wired solutions.
In contrast to their wired counterpart, wireless connectivity offers scalability, flexibility, and cost efficiency because it reduces the need for expensive wires and installation costs.
%
%

Today, wired networks are popular because the microgrid and energy system domains set special requirements for communications, such as latency, reliability, availability, and throughput.
For example, failures or anomalies in data delivery can cause severe damages in the control and automation used in microgrids. 

\subsection{State of the Art}
The state-of-the-art research includes many studies which aim to address the performance evaluation of specialized protocols, such as IEC-61850 over a private LTE network described in \cite{eriksson2017performance}. Here, a prototype emulates an \ac{IED}, which was derived from an IEC-61850-90-5 open-source project.
In this work, different metrics, such as reliability, availability, latency, and throughput, were evaluated. 
Based on a simulation analysis, the authors of \cite{mms_lte_karaginannis} investigated whether \ac{LTE} can be used in combination with IEC-61850 and \ac{MMS} to support smart metering and remote control communication.
In \cite{musaddiq2014performance}, a substation automation system was modeled and simulated under a wireless and hybrid network for three types of message, viz. \ac{GOOSE}, \ac{SV}, and interbay trip messages. The authors conclude that when circuit breakers were connected wirelessly, trip messages had high delays.
Many other works provide similar performance evaluations (\cite{kheaksong2016performance};  \cite{parikh2010opportunities}). However, most of them only address ideal scenarios using simulations. 

In contrast to the current state of the art, the results reported in this chapter are considered fully realistic scenarios, which are represented by real rural microgrid users and a practical deployment of a private cellular network that is based on a \ac{5G} architecture that involves the application of edge computing to reduce the backhaul latency.
It also considers true radio propagation with a diversity of wireless networking impairments.
This condition makes the output of this work unique.

\section{Test Bed Description}
\label{S:1}
The main elements of the test bed are composed of the wireless communication network and the \ac{LVDC} microgrid pilot.

\subsection{Wireless Communication }
The connectivity used in the field test bed is based on a \ac{5G} architecture, which is mostly characterized by the utilization of an edge-computing implementation of the core network to avoid latency impairments between the \ac{RAN} and the core network, which is managed remotely through the cloud. 
The \ac{RAN} is based on \ac{LTE}-Advanced, which is operating in Band 40 with a bandwidth of 20 MHz.
%

%
The architecture of the wireless cellular network is based on an end-to-end private wireless network with an edge-computing platform. It aims to support real-time applications for industrial applications.
Compared with a traditional 4G cellular network, the architecture of the network deployed in the field has an edge server, where the main core network server elements are deployed. 
This edge-computing core network is capable of being configured remotely from anywhere on the Internet. However, as it is the edge, the latency constraints in the core network are avoided. 
Thus, only the wireless access network represents the most important source of degradation when the latency is the \ac{KPI}. 

The \ac{CPE} is compliant with the downlink category 15 and the uplink category 13 of the \ac{3GPP} specifications. 
Hence, the radio supports downlink carrier aggregation with 256 QAM and 2x2 \ac{MIMO} technology, and the uplink single carrier with 64 QAM 2x2 \ac{MIMO}.
Each \ac{CPE} has a fixed beam antenna with up to a 12.5 dBi gain and a power transmission of 23 dBm.
The base station operates with a power transmission of 35 dBm with an omnidirectional antenna with a gain of 4 dBi.

\subsection{LVDC Microgrid Pilot}
\label{S:2}
The \ac{LVDC} microgrid field pilot is part of a local distribution network owned by a Finnish energy company Suur-Savon Sähkö Oy (SSS Oy) and operated by its subsidiary distribution company Järvi-Suomen Energia Oy. The platform has been implemented in collaboration with Lappeenranta University of Technology (LUT) as part of the Finnish national Smart Grids and Energy Markets research program.

The system is designed to meet flexible testing needs and has exhaustive distribution system functionality. The pilot is implemented according to the national low voltage standard series SFS 6000 based on HD 60364, IEC 60364, and IEC 60664. The voltage quality requirements were set based on the EN 50160 standard, but some limit values were changed as reported in \cite{6683522}.\\
\subsubsection{Architecture}
\begin{figure}[!h]
    \centering
    \includegraphics[scale=.5]{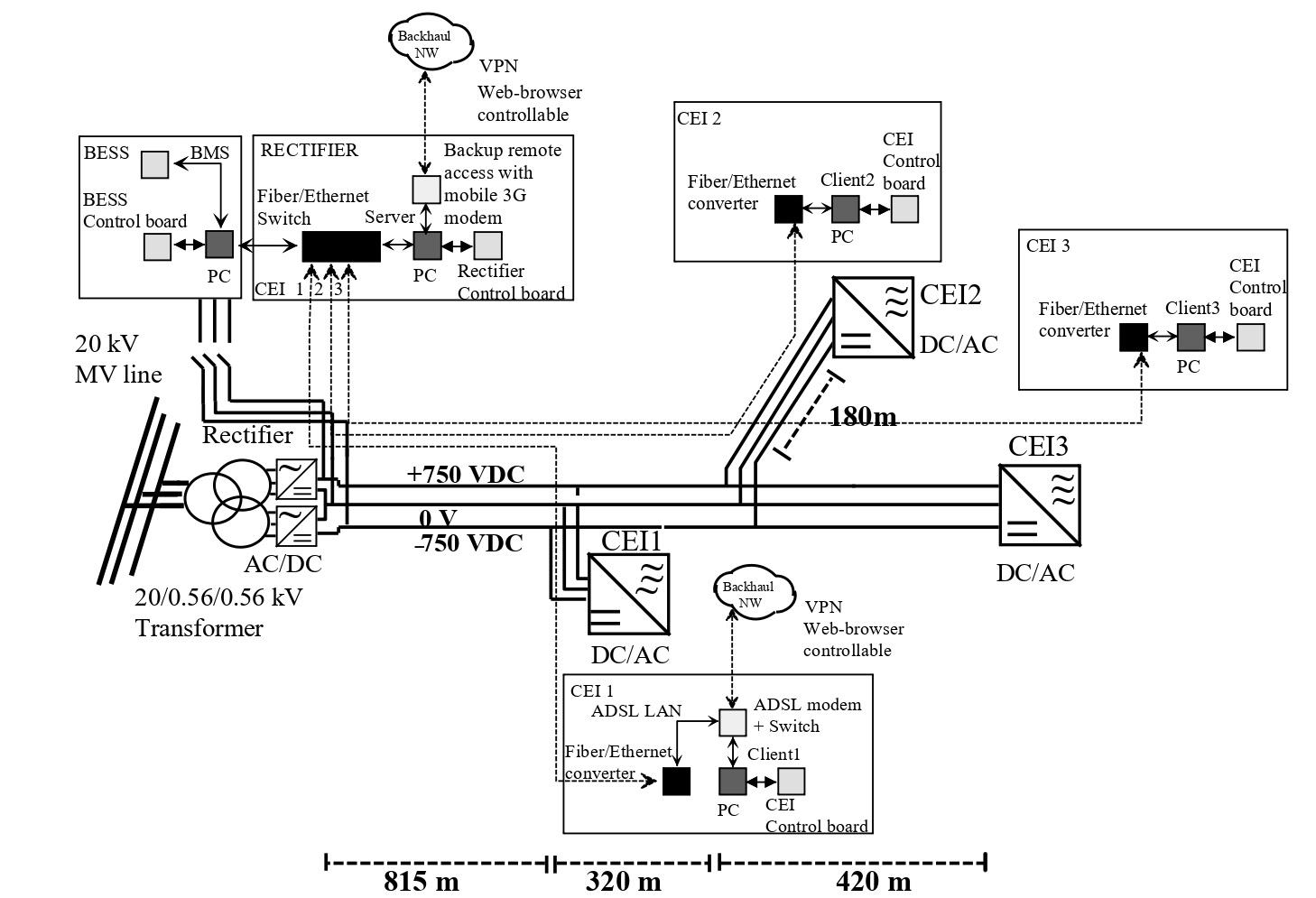}
    \caption{Grid configuration.}
    \label{fig:grid_config}
\end{figure}

The \ac{LVDC} pilot has a structure following the concept of the \ac{LVDC} distribution system. Figure~\ref{fig:grid_config} presents the grid configuration of the pilot. The setup consists of a 70 kVA-rectifying substation, supplied with a double-tier transformer from a 20 kV MV network, a \ac{BESS}, comprising 2 x 30 kWh-nominal capacity sections, 2 x 35 kVA converters and a 100 kVA transformer, a 1.7 km-long underground-cabled bipolar ±750 V DC network, three 16 kVA \acp{CEI}, and a \ac{PV} array, which is currently out of operation. Two CEIs are connected to the plus pole of the ±750 V DC network, and one CEI is connected to the negative. The CEIs supply four end consumers with three-phase 230/400 V AC  voltage. The third CEI supplies two end users as their consumption is lower than the nominal power of the inverter. CEIs have a three-phase structure with galvanic isolation that allows the use of terrain-isolated functionalities of unearthed (IT) grounding in the DC distribution grid with the functionality of earthed (TN) grounding of the end consumer, which is typically used in Finland: for further details, the reader is referred to \cite{6777363}.\\
The system is normally operated in the grid connection mode. However, in case of MV supply interruption, it is switched into the island operation mode, where \ac{CEI}s are supplied from the \ac{BESS}. The voltage variation in the DC grid is kept within +10\% -20\% during normal operation and -30\% in the isolated operation mode. The end consumers have 50±0.2\% of frequency variation, ±1\% in normal operation and ±15\% in island mode operation; voltage variations are used according to \cite{6302501}. 

\subsubsection{Application}
\begin{figure}[!h]
    \centering
    \includegraphics[scale=.6]{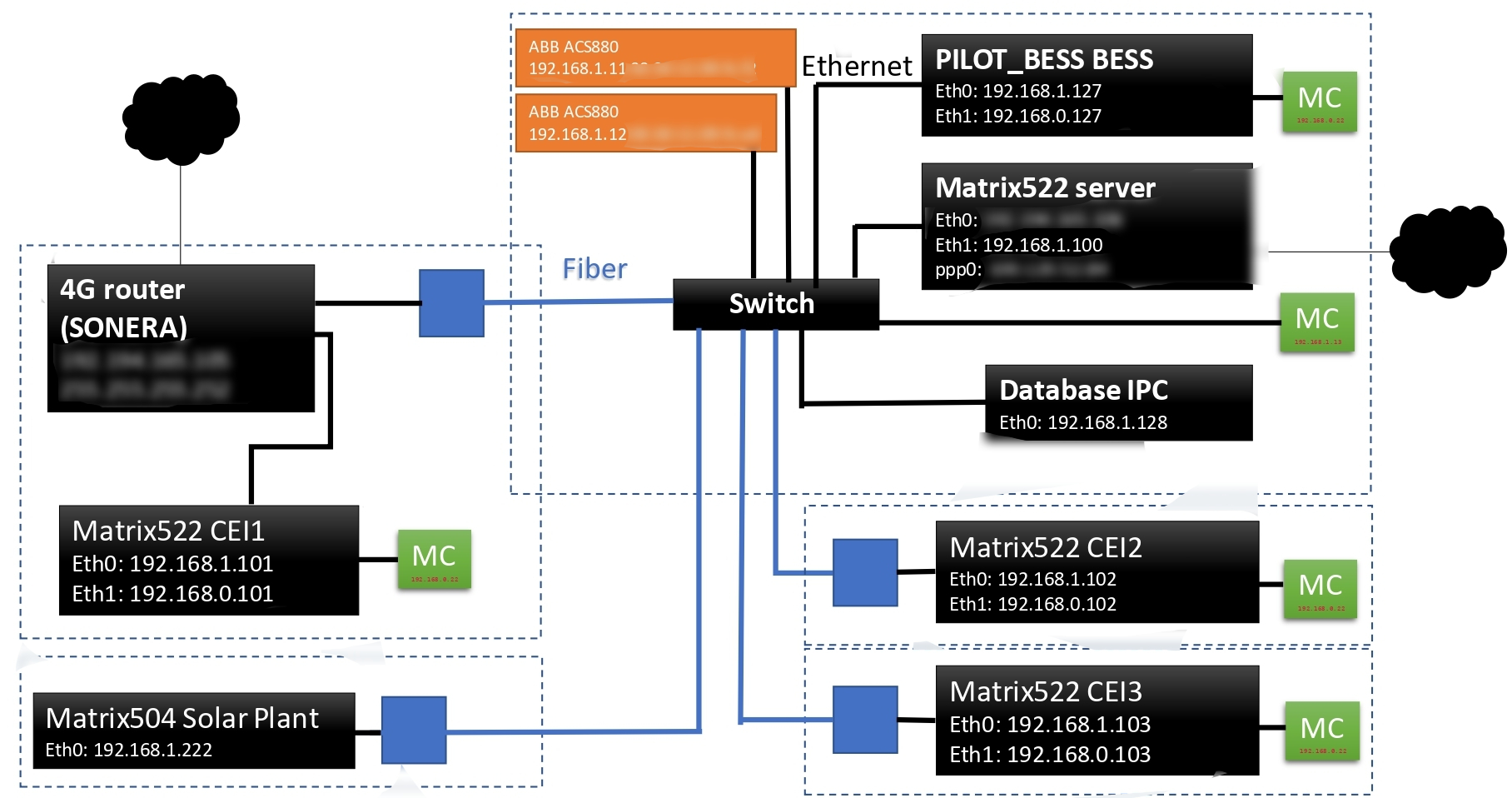}
    \caption{Wired network structure.}
    \label{fig:Network_configuration_init}
\end{figure}

\begin{figure}[hbt!]
    \centering
    \includegraphics[height=8.5cm]{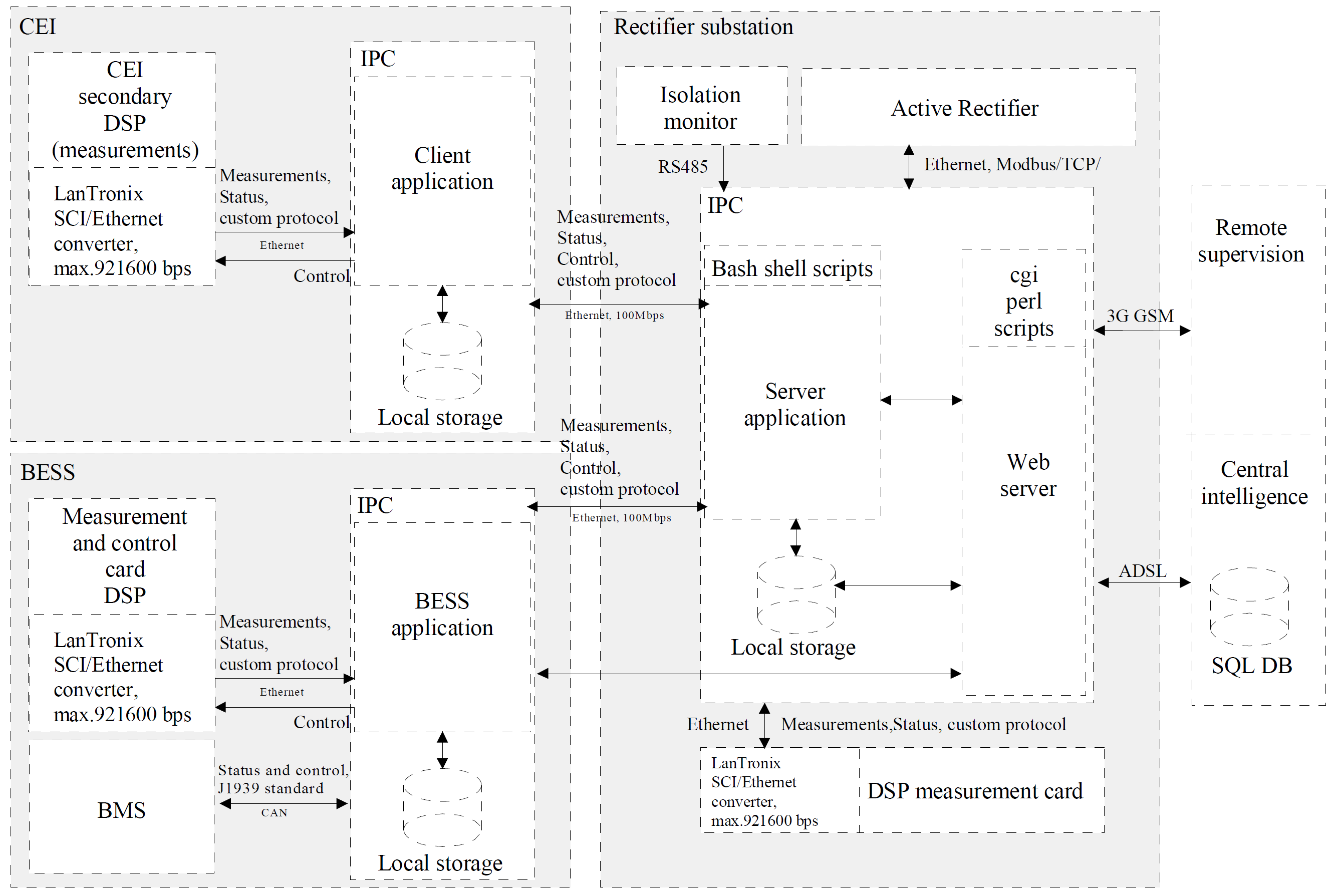}
    \caption{ICT architecture.}
    \label{fig:Application_architecture}
\end{figure}

Figure~\ref{fig:Network_configuration_init} illustrates the network structure with a wired \ac{FO} communication network, installed alongside  \ac{LVDC} underground cables. \ac{FO} is a single-mode multicore fibre with 16 cores. \ac{CEI}s are connected with a rectifier with a core per connection. The information and communication system of the pilot is implemented with a Linux-based \ac{IPC} and has a client-server architecture. Each \ac{CEI} has an \ac{IPC}: the model is Matrix 522, which is connected with a measurement card and applies monitoring, control, data logging, and web--client functions. The solar plant is equipped with another model of \ac{IPC} (Matrix504), which provides functionality similar to that of \ac{CEI} devices. The BESS control and monitoring, server, and database are implemented on an \ac{IPC} placed at the rectifier substation (\cite{7152001}). The server has the following functionalities: data gathering from all client applications, gateway of remote connection, updates of the web portal, data logging, control of CEIs, and data transfer to the remote database. The collected data are also stored in the local database, located at the SSD connected to the database \ac{IPC}.
Figure~\ref{fig:Application_architecture} illustrates the architecture of the \ac{ICT} system of the pilot, showing the monitoring and control solutions. The web portal provides remote control and uninterrupted monitoring of the system. All supervised data are presented on the portal and logged and stored locally. The web portal allows to monitor the following operation monitoring functions ( \cite{7152001}):
\begin{itemize}
    \item Statuses of the network and its components
    \item Power flow monitoring---the data consist of minimum, maximum, and average power, voltage and current values with one-minute resolution.
    \item Power quality monitoring---includes harmonic content, defined by the fast-Fourier transform (FFT) and power quality indices. The measurement intervals are 16 and 18 s for the CEIs and the rectifier, respectively. 
    \item System supervision portal---allows access to the pilot without installation of additional software. The update rate of the web portal is 10 s. 
    \item Fault identification---records the fault code and the corresponding measurements window of current and voltages with time stamps and notifies the support personnel. The measurements of currents and voltages are applied at the rectifier substation.  
    \item Network rectifier station---controls and monitors over a network front-end rectifier, which is required to ensure reliable network operation. 
    \item \ac{BESS} monitoring and control---implements full control of the \ac{BESS} and condition monitoring. It allows automatic transfer from the island mode and normal operation. 
    \item CEI control diagnostic---is designed to meet the requirements of the public network, actual end users, and research. 
\end{itemize}

\subsection{Test Bed Objectives}
The main objective of this test bed is to benchmark and evaluate the performance of a real private cellular network located in the rural area of Suomenniemi (southern Finland), where a microgrid pilot provides power supply to local citizens.
The objectives of this study are the following:

\begin{itemize}
    \item Evaluate the system performance of a cellular network that is based on a \ac{5G} architecture using an edge core network to reduce potential impairments on the backhaul between the radio and core networks.
    \item Evaluate the throughput between \acp{IED} and an edge cloud device. Each \ac{IED} is connected to the cellular network \ac{CPE}, and the edge cloud device is connected to the \ac{PDN} gateway in the \ac{LTE}--\ac{CPE} (core network).
    \item Evaluate the system performance in terms of throughput and latency in different scenarios, such as data rate, simultaneous traces for \ac{TCP}, and \ac{UDP} packages.
    \item Evaluate the performance of the standard IEC-61850 based on the premise that the protocols, such as \ac{MMS}, \ac{GOOSE}, and \ac{SV}, are enabled by \ac{TCP} and \ac{UDP} on a wireless network.
    \item Generate a radio coverage mapping based on measurements obtained in the field and a temporal analysis of typical LTE metrics, such as \ac{RSSI}, \ac{RSRP}, \ac{RSRP}, and \ac{SINR} in each point where the \ac{LTE} \acp{CPE} are installed.
\end{itemize}

\section{IEC-61850}
The preparation of IEC 61850 began in 1995 in the United States. Initially, the work was done by two independent, parallel working groups: The first one is the UCA International Users Group, developing a common object model of substation equipment (GOMFSE); the second was formed within the framework of the IEC Technical Committee 57 and was engaged in the creation of a standard of data transmission protocols at a substation. In 1997, the work of both groups was combined under the auspices of Working Group 10 of IEC TC 57 and became the basis of the IEC 61850 standard.\\
The standard is based on the following statements:
\begin{itemize}
    \item Technological independence, i.e., regardless of technological progress, the standard should be subject to minimal changes over time;
    \item Flexibility, i.e., it should allow for different tasks to be solved using the same standardized mechanisms;
    \item Extensibility, i.e., meeting the requirements, the standard allows  to meet the changing needs of the electric power industry and use the latest advances in computer, communication, and measurement technologies.
\end{itemize}
The main features of IEC-61850:
\begin{itemize}
    \item It defines not only how, but also what information should be exchanged. The standard describes abstract models of the object's hardware and the functions performed. The information model underlying the standard is presented in the form of classes of data objects, data attributes, abstract services, and descriptions of the relationships between them;
    \item It defines the system design and setup process;
    \item It defines the system configuration description language (SCL). This language provides the ability to exchange information about the configuration of devices in a standardized format between software from different manufacturers;
    \item It describes the methods of testing and acceptance of equipment.
\end{itemize}
The key aspect of IEC-61850 defined before is the definition of a three-level automation system architecture, which comprises the following: the process level, the bay level, and the station level.
The process level includes sensors, remote I/Os, and actuators. The bay level consists of control, protection, and monitoring intelligent electronic devices (IED). The station level comprises the station computer with databases, the operators' user interfaces (UI), and remote communication. 
The standard has an object-oriented approach that allows to establish logical communication among IED, a merging unit (MU), and the system operator. \\
Initially, the standard defined only an automation system within a single object, and the links between substations were not included in the model. In 2012, IEC-61850 was expanded with parts 90-1 and 90-5. The updated model presented the communication system architecture described in the second edition of the standard, which also provides connections between substations. The structure of information exchange is described between the levels, as well as within each of them. These parts allow the expansion of the communication standardization beyond substation limits and make it possible to apply the standard also to smart grids.\\
IEC-61850 defines three key communication protocols to implement data exchange within and between defined levels: 
\begin{itemize}
    \item \acs{MMS} (Manufacturing Message Specification) is a client-server \acs{TCP} protocol using real-time data exchange among network devices and computer applications. It is described in Part 8-1 of IEC-61850.
    \item \acs{GOOSE} (Generic Object Oriented Substation Events) is a publisher--subscriber \acs{UDP} protocol applied for ultra-high messaging speed transmission of grouped status or value data and to ensure enhanced reliability. It is also described in Part 8-1 of IEC-61850.
    \item \acs{SV} (Sampled Values) is a publisher--subscriber \acs{UDP} protocol, used for data exchange between an MU and an IED with an exactly defined time interval. It is described in Parts 9-2 and 9-2LE of IEC-61850.
\end{itemize}
\subsection{MMS}
Part 8-1 of IEC 61850 describes the \acs{MMS} (Manufacturing Message Specification) protocols used for data exchange for which time delay is not critical. They operate on the full OSI model on top of the \acs{TCP} stack, which means that data transmission over this protocol is carried out with relatively large time delays. For example, this protocol can be used to send remote control commands, collect telemeasurement and telesignaling data, and send reports and logs from remote devices. The detailed presentation of the protocol can be found in \cite{Typhoon_mms}. 
The \acs{MMS} defines the following:
\begin{itemize}
    \item A set of standard objects on which operations are performed that must be present in the device (e.g., reading and writing variables, signaling events);
    \item A set of standard messages that are exchanged between the client and the server for control operations;
    \item A set of rules for encoding these messages (how values and parameters are assigned to the corresponding bits and bytes when forwarding);
    \item A set of protocols (rules for exchanging messages between devices).
\end{itemize}
Therefore, the \acs{MMS} protocol itself is not a communication protocol: it only defines the messages that should be transmitted over a specific network. As a communication protocol, \acs{MMS} uses the \acs{TCP}/IP stack, and thus, it belongs to the transport layer of the OSI model.
\subsection{GOOSE}
Part 8-1 also describes the \acs{GOOSE} protocol as a \acs{UDP} publisher/subscriber-type communication. \cite{Typhoon_goose} describes the details of the protocol and its application. The information content of this protocol is defined in IEC 61850-7-2. There is a description of the protocol syntax, the definition of the data assignment of an Ethernet frame in accordance with ISO/IEC 8802-3, and the procedures related to the use of this frame. Because the data in this protocol are assigned directly to the Ethernet frame, bypassing the \acs{TCP} stack, the data transmission in it is carried out with significantly lower time delays compared with \acs{MMS}. This allows the \acs{GOOSE} protocol to be used to transmit commands to disconnect the circuit breaker from the protection and other signals for which the time delay is critical.
Initially, the data sets are grouped for their subsequent sending using the \acs{GOOSE} protocols. Then, in the \acs{GOOSE}-sending control unit, a reference to the created dataset is specified: in this case, the publisher knows exactly what data to send. Within the framework of a single \acs{GOOSE} message, both one value (e.g., the overcurrent protection trigger signal) and several values can be sent simultaneously (e.g., the start and overcurrent protection trigger signals). The subscriber can extract from the packet only the data that it needs.
The transmitted \acs{GOOSE} message packet contains all the current values of the data attributes entered in the dataset. If any of the attribute values change, the device immediately initiates sending a new \acs{GOOSE} message with the updated data.


\subsection{SV Messages}

\begin{figure}[hbt!]
    \centering
    \includegraphics[height=8.5cm]{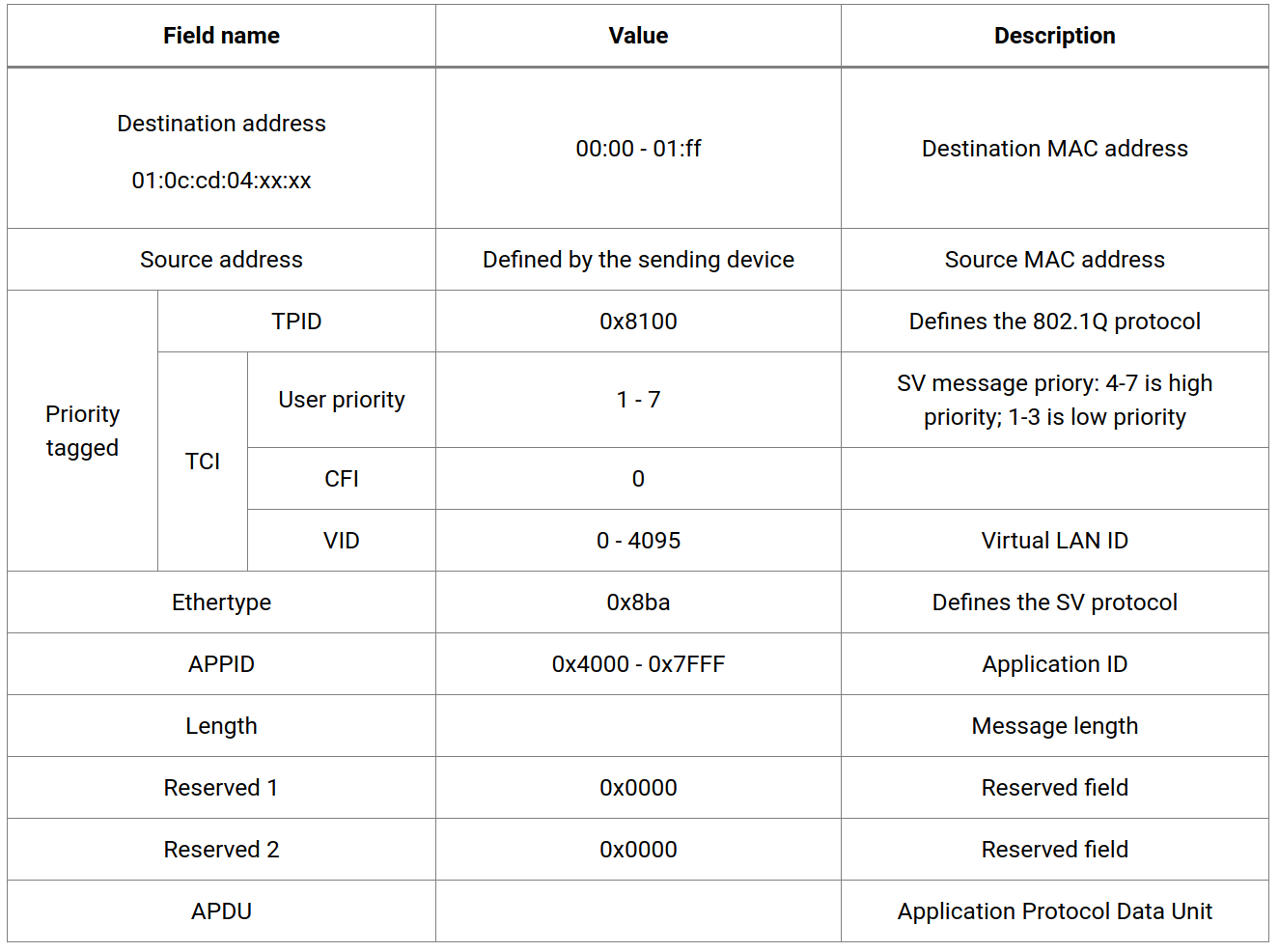}
    \caption{Structure of an SV message.}
    \label{fig:SV_message_structure}
\end{figure}

The \acs{SV} protocol, defined in IEC-61850 9-2, is used to transmit measured values from the process to the bay level with a predefined time interval within a power facility. The protocol is described in detail in \cite{Typhoon_sv}. The data transmission is critical to the latency and should have a high bandwidth. The time interval depends on the measured signal frequency and Samples Per Period (SPP). Part 9-2LE of IEC618950 defines two possible SPPs in 80 and 256 samples. Therefore, for example, for the frequency of 50 Hz, the time interval will be equal to $1/(80\cdot 50)=250 \mu s$. The maximum delay for \acs{SV} multicast data transmission is $< 208.3 msecs$ (\cite{CiscoSV}).
 Figure~\ref{fig:SV_message_structure} illustrates the structure of an \acs{SV} message. The APDU field contains the payload of the \acs{SV} messages. Every APDU (Application Protocol Data Unit) contains up to eight ASDUs (Application Specific Data Units), where each ASDU has a unique \acs{SV} identification value and contains one three-phase current and voltage measurements. 
 \acs{SV} is a \acs{UDP} protocol which has publisher/subscriber structure. All messages are published under a topic. A subscriber receives all \acs{SV} messages from all publishers, and then filters and parses according to the subscribed topic. \\
\subsection{R-SV, R-GOOSE}
\acs{SV} and \acs{GOOSE} protocols are defined in IEC-61850 9-2 and 8-1, respectively. Both operate in a publisher/subscriber mode over layer 2 in a generic network protocol stack. As a consequence, the data transmission is limited to \acp{LAN}.
There are two options to transmit \acs{GOOSE} or SV messages over a \ac{WAN} (\cite{8979307}). The first one is tunneling messages over a high-speed network, such as SDH or SONNET. 
The second option is application of Routable-SV/GOOSE (R-SV/R-GOOSE) defined in the  extended version of the protocols (IEC-61850-90-5) by adding a network and transport layer. The first option is less preferable as it leads to employment of a gateway, and message exchange is carried out in a point to point manner. R-SV messages, in turn, can be multicasted in WANs. The R-SV protocol also includes a Group Domain of Interpretation (GDOI) security mechanism for protecting data transmission  \cite{8979307}.

\subsection{Summary of the Standard IEC-61850}

\begin{figure}[hbt!]
    \centering
    \includegraphics[height=6cm]{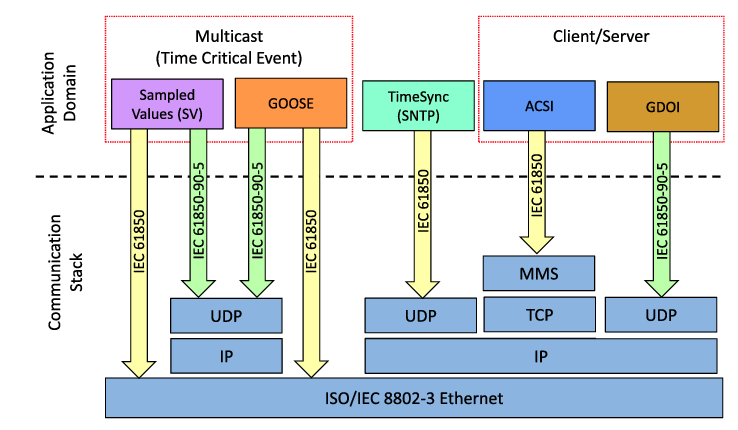}
    \caption{Structure of the protocol application.}
    \label{fig:protocol_strcture}
\end{figure}

IEC 61850 is an Ethernet-based international standard, which has become the standard for communication at many energy facilities and substations. The IEC 61850 standard is aimed to integrate all protection, control, measurement, and monitoring functions into the energy facilities. The control monitoring and protection functions are placed at the IED: the measurement one is applied with an MU. Figure \ref{fig:protocol_strcture} illustrates the structure of the protocol application. 
Figure \ref{fig:protocol_strcture} illustrates the structure of the protocol application. \acs{MMS} is a \acs{TCP} client/server-based protocol used for non-time-critical control signals. \acs{GOOSE} is a \acs{UDP} publisher/subscriber-based protocol applied for the transmission of time-critical control and protection-related data. \acs{SV} is a \acs{UDP} publisher/subscriber-based protocol utilized, for instance, for measurement data exchange with a constant sampling frequency. The latency and bandwidth are critical for \acs{SV} data transmission. \acs{GOOSE} and \acs{SV} can be used only within LAN. R-GOOSE and R-SV are extension versions of the protocols that allow transmission of data over WAN. Therefore, IEC-61850 can be used not only within substations but also for smart grids. \acs{MMS} is a \acs{TCP} client/server-based protocol used for non-time-critical control signals; \acs{GOOSE} is a \acs{UDP} publisher/subscriber-based protocol applied for the transmission of time-critical control and protection-related data; and \acs{SV} is a \acs{UDP} publisher/subscriber-based protocol used for measurement data exchange with a constant sampling frequency. The latency and bandwidth are critical for \acs{SV} data transmission. Furthermore, \acs{GOOSE} and \acs{SV} can be used only within LAN. R-GOOSE and R-SV are extension versions of the protocols that allow transmission of data over WAN. Therefore, IEC-61850 can be used not only within substations but also for smart grids \cite{summary_picture}.

\section{Test Bed Setup and Tools}
\label{S:3}

\subsection{Deployment}

\begin{figure}[hbt!]
    \centering
    \includegraphics[height=7cm]{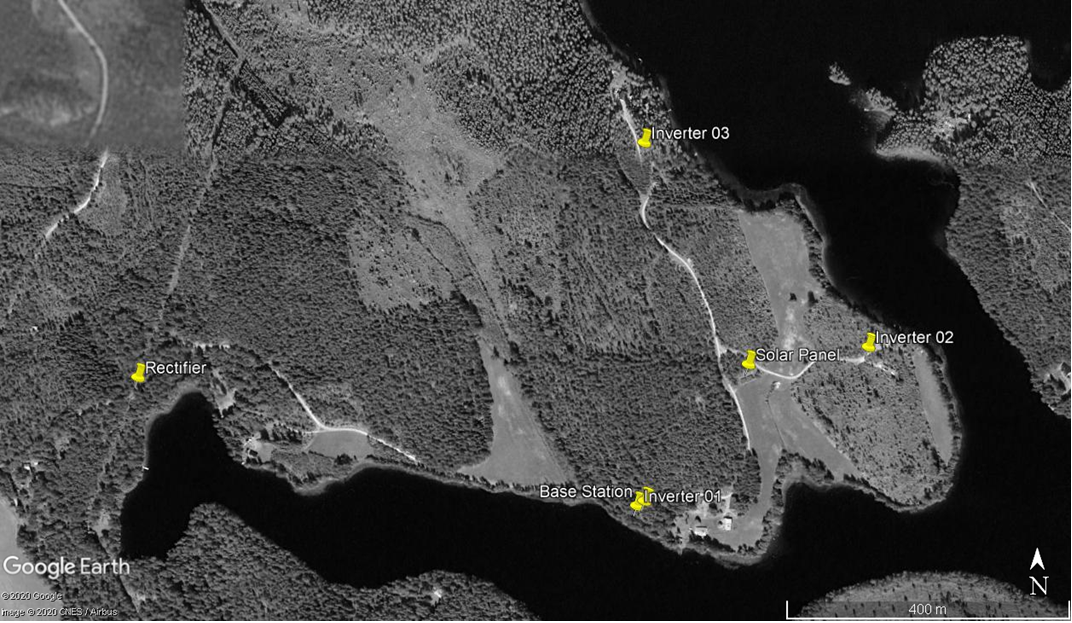}
    \caption{Geographical location of the microgrid elements. A \ac{CPE} is installed in each of these points.}
    \label{fig:generic_view}
\end{figure}

The microgrid is running in Suomenniemi (southern Finland), which is a small community with a high forest density and some challenging elevations that generate impairments in the radio propagation. 
Figure \ref{fig:generic_view} visualizes the elements of the network. In each microgrid element, one \ac{CPE} is installed.

\begin{figure}[hbt!]
    \centering
    \includegraphics[height=7cm]{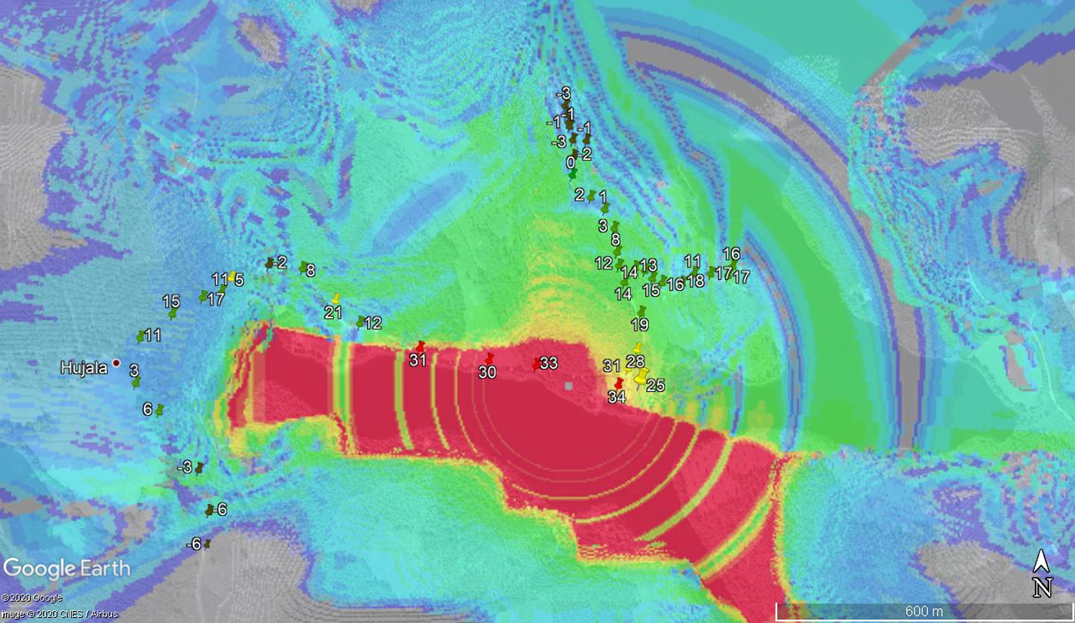}
    \caption{Radio propagation coverage with some specific \ac{SINR} values in some points obtained during the drive test. Red color indicates a high \ac{SINR}, green a medium \ac{SINR}, and blue the cellular edge border (SINR$<$0 dB).}
    \label{fig:coverage_prediction}
\end{figure}

During the deployment, a preanalysis was conducted based on a radio propagation prediction of the network coverage.
This radio coverage prediction was made using a simplified link budget that depends of the power transmission, antenna gain, and height location of antennas. 
The output of this analysis is presented in Figure \ref{fig:coverage_prediction}.
In the case of the base station, the antennas were installed above 25 m and the \acp{CPE} to an average height of 3 m.
%

\begin{figure}[!h]
    \centering
    \includegraphics[height=8cm]{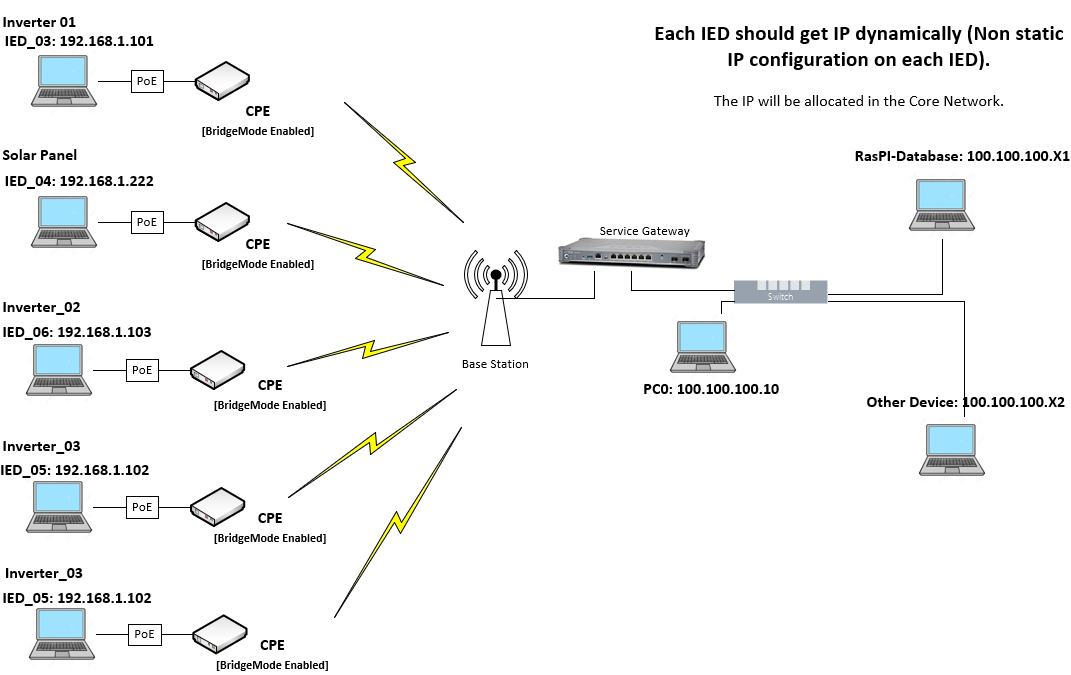}
    \caption{Generic architecture of the network used during the system performance evaluation in the field.}
    \label{fig:architecture}
\end{figure}

The test setup is composed of a base station, which is compliant with the LTE-advanced specifications, an \ac{EPC}, which is compliant with \ac{5G} architectures using edge computing specifications, and five \acp{CPE}, which are the capillary access of the network.
One computer, named PC0, is connected to the core routing unit, and each \ac{CPE} is connected to one embedded system, as shown in Figure \ref{fig:architecture}.
To avoid routing issues on the network, the bridge mode was enabled in each \ac{CPE}.

The tool used to evaluate the throughput is IPERF \cite{dugan2010iperf}. There are many testing configuration setups that are evaluated using an automatic Python script, which was used to manage the measurements to be made in the network.
The scripts used in this stage are publicly available in the following repository: 
\\
\textbf{https://github.com/aikonbrasil/NetMeter}.

\subsubsection{Database Description}
\begin{figure}[hbt!]
    \centering
    \includegraphics[height=5cm]{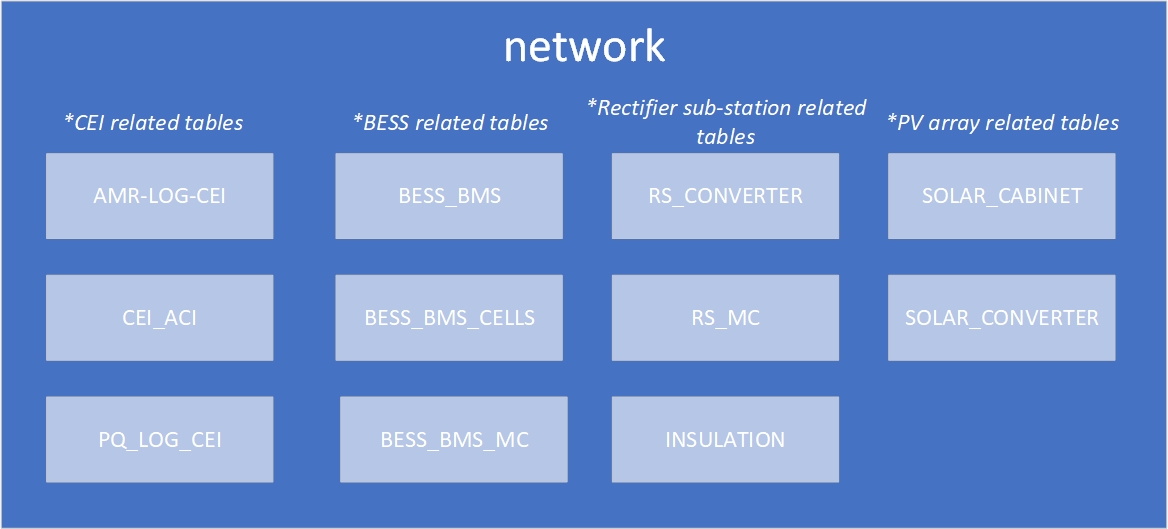}
    \caption{Generic database structure of the \ac{LVDC} system used in the field.}
    \label{fig:database_str}
\end{figure}

Figure~\ref{fig:database_str} illustrates the structure of the local database where data gathered from all applications are stored. The database is placed on the interconnection of SSD and Raspberry Pi 4 and located at the rectifier station. It contains eleven tables, where each table includes data related to the specific \ac{LVDC} grid application. The database is used in the testing as a source for creating emulated data packages. As the \ac{RS}, the \ac{BESS}, and the database are located at the same place, the data from the corresponding tables are not used in testing as the \acp{IPC} of these devices and the database have a wired connection that is not replaced with a wireless system. \ac{CEI}s and solar installations are under a wireless connection through \acp{CPE}; therefore, the data from the \acp{CEI} and the solar applications can be used for testing. However, because the PV array was out of operation at the moment of testing and only a small part of the information was available from that installation, only data from \ac{CEI}-related tables were used in testing.\\

There are four groups of tables related to the different equipment: \ac{CEI}, \ac{BESS}, \ac{RS}, and PV. The \ac{BESS} group comprises three tables: BESS\_BMS, where data from the battery management system are collected (e.g., total voltage, total current, state of charge); BESS\_BMS\_CELLS, where data about cells are included (e.g., cell id, cell voltage, cell temperature); and BESS\_BMS\_MC, which contains information from the measurement card of the \ac{BESS} (e.g., \ac{BESS} grid voltage, \ac{BESS} grid current, \ac{BESS} power). Another group is related to the rectifier and comprises three tables: RS\_CONVERTER, where general data about the converter are gathered (e.g., power, line current, DC voltage); RS\_MC including information from the measurement card of the rectifier station (e.g., grid voltages and currents, FFT and THD of currents); and INSULATION table that comprises alarm and fault-related logs. As CEI tables are used in testing and PV-related tables can be used after returning PV into operation, a more detailed description of the attributes of the tables is presented in Tables~\ref{tb:CEI_tables_1}--\ref{tb:PV_tables}.
\begin{table}[hbt!]
	\caption{CEI table description part 1.}
	\label{tb:CEI_tables_1}
	\begin{center}
		\begin{tabular}{c|p{0.25\textwidth}|p{0.53\textwidth}}
			\hline
			\textbf{Table name} &  \textbf{Column name} & \textbf{Description} \\
			\hline
			ARM\_LOG\_CEI & TIMESTAMP & Time on the client side \\
			\hline
			ARM\_LOG\_CEI & CEI\_ID & ID of the client (from which inverter data are sent) \\
			\hline
			ARM\_LOG\_CEI & U\_DC & Input inverter voltage (before inverting), direct voltage\\
			\hline
			ARM\_LOG\_CEI & I\_DC & Input inverter current (before inverting), direct current \\
			\hline
			ARM\_LOG\_CEI & U\_RMS\_A, U\_RMS\_B, U\_RMS\_C & Phase voltages on the inverter output side (3-phase inverter), alternate voltages (AC) \\
            \hline
			ARM\_LOG\_CEI & I\_RMS\_A, I\_RMS\_B, I\_RMS\_C & Phase currents on the inverter output side (3-phase inverter), currents voltages (AC) \\
			\hline
			ARM\_LOG\_CEI & S\_RMS\_A, S\_RMS\_B, S\_RMS\_C & Phase powers on the inverter output side (3-phase inverter), phase powers (AC) \\
			\hline
			ARM\_LOG\_CEI & T\_IGBT & Temperature at the switches (internal inverter sensor) \\
			\hline
			ARM\_LOG\_CEI & T\_AMBIENT & Temperature in the cabinet  \\
			\hline
			ARM\_LOG\_CEI & STATUS\_FLAGS & Based on this status, THD values are set to 0 or factual values. \\
			\hline
            ARM\_LOG\_CEI & FAULT\_FLAGS & Flag shows if there are any faults on the inverter side \\
			\hline
		\end{tabular}
	\end{center}
\end{table}
\begin{table}[hbt!]
	\caption{CEI table description part 2}
	\label{tb:CEI_tables_2}
	\begin{center}
	\begin{tabularx}{\textwidth}{X|p{0.25\textwidth}|p{0.55\textwidth}}
		\hline
			\textbf{Table name} &  \textbf{Column name} & \textbf{Description} \\
		    \hline
		    CEI\_ACI & TIMESTAMP & Time on the client side \\
			\hline
			CEI\_ACI & CEI\_ID & ID of the client (from which inverter data are sent) \\
			\hline
			CEI\_ACI & T\_L1, T\_L2, T\_W1, T\_W2 & ACI box temperatures \\
			\hline
			CEI\_ACI & R1-R6 & States of relays in the inverter ACI box \\
			\hline
			CEI\_ACI & Connected & Connection state \\
			\hline
			PQ\_LOG\_CEI & TIMESTAMP & Time on the client side \\
			\hline
			PQ\_LOG\_CEI & CEI\_ID & ID of the client (from which inverter data are sent) \\
			\hline

		\end{tabularx}
	\end{center}
\end{table}
\begin{table}[hbt!]
	\caption{PV table description.}
	\label{tb:PV_tables}
	\begin{center}
		\begin{tabularx}{\textwidth}{c|X|p{0.5\textwidth}}
			\hline
			\textbf{Table name} &  \textbf{Column name} & \textbf{Description} \\
			\hline
			SOLAR\_CABINET & TIMESTAMP & Time on the client side \\
			\hline
			SOLAR\_CABINET & CABINET\_ID  & ID of the solar plant (there is only one in the system) \\
			\hline
			SOLAR\_CABINET & Heater, fan   & States of heating or cooling operation device \\
			\hline
			SOLAR\_CABINET & CB & Current breaker state \\
			\hline
			SOLAR\_CABINET & Tcabinet  & Temperature in the solar cabinet \\
			\hline
			SOLAR\_CABINET & Connected  & Connection state \\
			\hline
			SOLAR\_CONVERTER & TIMESTAMP & Time on the client side \\
			\hline
			SOLAR\_CONVERTER & CABINET\_ID  & ID of the solar plant (there is only one in the system) \\
			\hline
			SOLAR\_CONVERTER & DCUPOWER  & Ensto DC/DC converter internal power reference \\
			\hline
			SOLAR\_CONVERTER & ActualPowerReference  & actual power reference of the solar plant \\
			\hline
			SOLAR\_CONVERTER & DC\_Link\_Current  & Current in the DC link between the grid and the solar plant \\
			\hline
			SOLAR\_CONVERTER & DC\_Link\_Power  & Power in the DC link between the grid and the solar plant \\
			\hline
			SOLAR\_CONVERTER & DC\_Link\_Voltage  & Voltage in the DC link between the grid and the solar plant \\
			\hline
			SOLAR\_CONVERTER & DC\_Out\_Current  & Solar output current \\
			\hline
			SOLAR\_CONVERTER & DC\_Out\_Power  & Solar output power \\
			\hline
			SOLAR\_CONVERTER & DC\_Out\_Voltage  & Solar output voltage \\
			\hline
			SOLAR\_CONVERTER & Temp1 ,Temp2  & Temperatures in a converter \\
			\hline
			SOLAR\_CONVERTER & SetpointPowerReference  & Set power reference for the solar plant \\
			\hline
			SOLAR\_CONVERTER & MSW  & Main status word \\
			\hline
			SOLAR\_CONVERTER & Fan  & Cooling state \\
			\hline
			SOLAR\_CONVERTER & Connected  & Connection state  \\
			\hline

		\end{tabularx}
	\end{center}
\end{table}


\section{Measurement Procedure and Results}
Three different procedures were applied in the measurements.
The first one is based on an isolated and simultaneous analysis of the \acs{TCP} and \acs{UDP} protocols between PC0 node and each of the embedded systems installed in each \ac{CPE}; this analysis is done in terms of throughput. 
The second procedure is based on the analysis of the latency evaluated with the \ac{ICMP}. 
Finally, the third procedure focuses on the analysis of the application. Thus, the database information was emulated in each embedded system, following the sampling time in which the data are generated in the real \acp{IED}. This information is transmitted via \ac{TCP} and \ac{UDP} to the embedded system that hosts the database.
In this particular procedure, the data integrity was evaluated on the received side to obtain numerical values indicating potential package loss. 

The results of each procedure are described in the following.

\subsection{Throughput Performance Evaluation} 
\begin{figure}[hbt!]
    \centering
    \includegraphics[height=7.8cm]{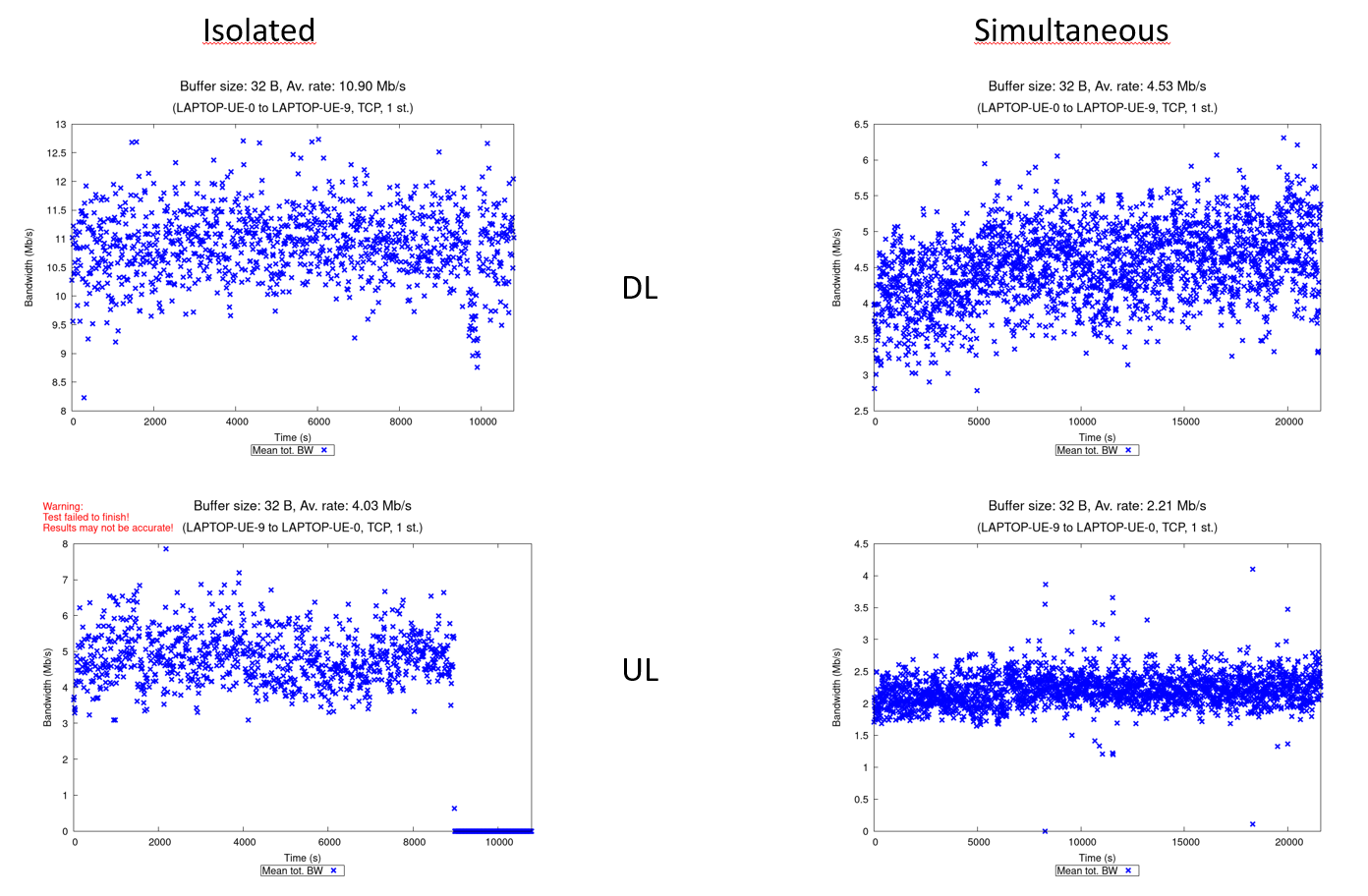}
    \caption{Performance evaluation carried out in the embedded system connected to the \ac{CPE} installed at the rectifier.}
    \label{fig:throughput_01}
\end{figure}

\begin{figure}[hbt!]
    \centering
    \includegraphics[height=7.5cm]{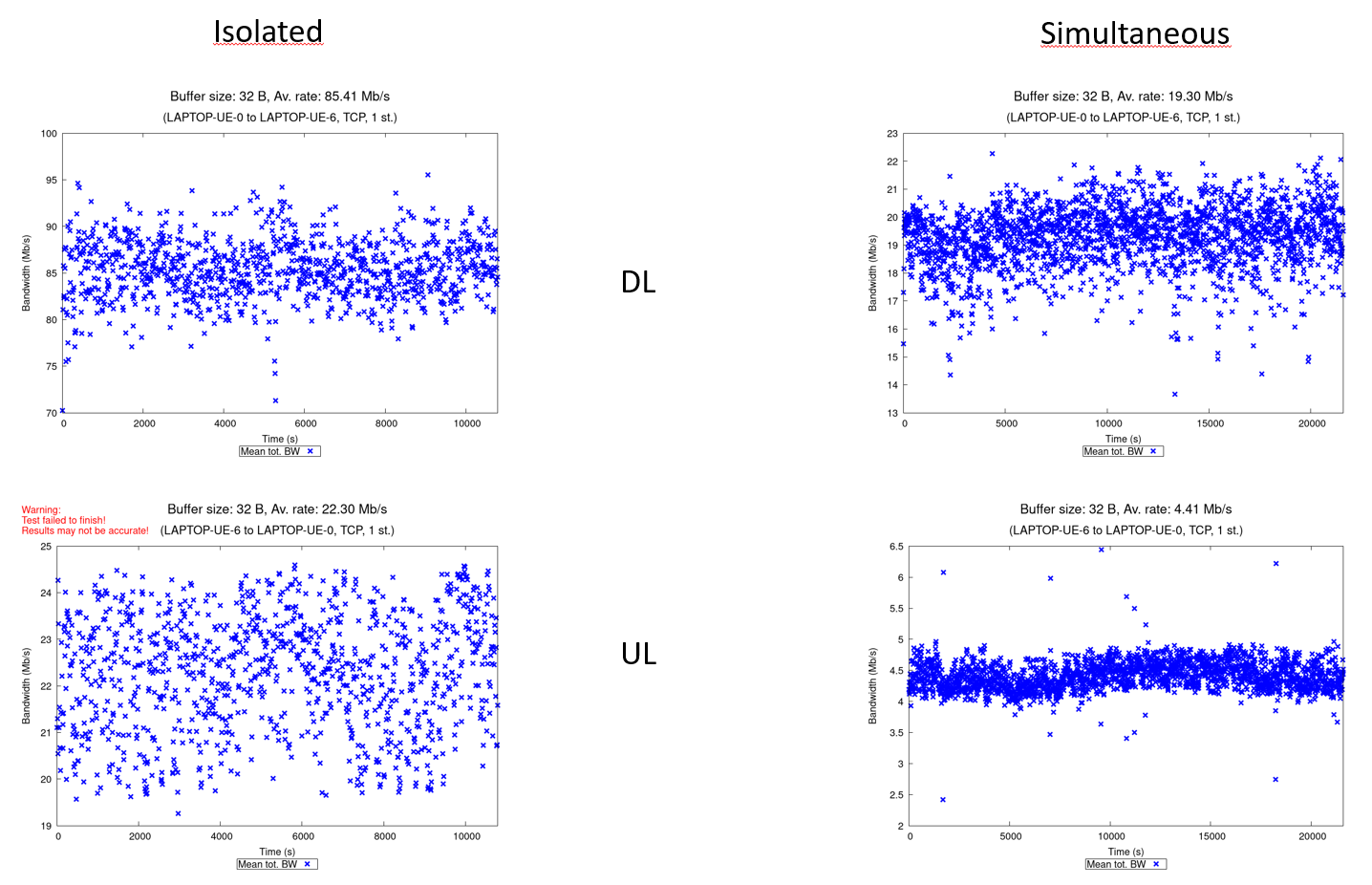}
    \caption{Performance evaluation carried out in the embedded system connected to the \ac{CPE} installed at Inverter 01.}
    \label{fig:throughput_02}
\end{figure}

\begin{figure}[hbt!]
    \centering
    \includegraphics[height=7.5cm]{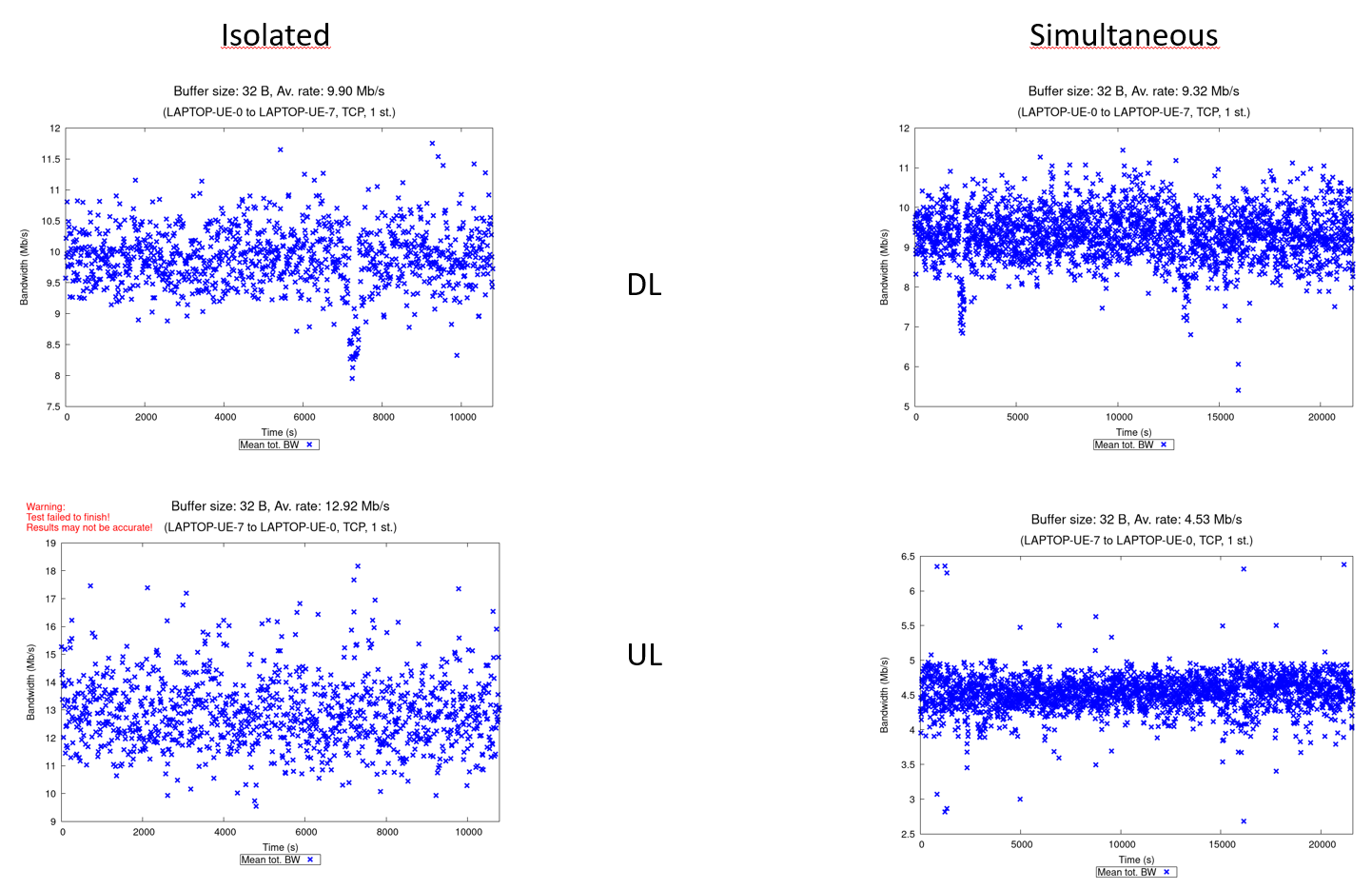}
    \caption{Performance evaluation carried out in the embedded system connected to the \ac{CPE} installed at Inverter 02.}
    \label{fig:throughput_03}
\end{figure}

\begin{figure}[hbt!]
    \centering
    \includegraphics[height=7.5cm]{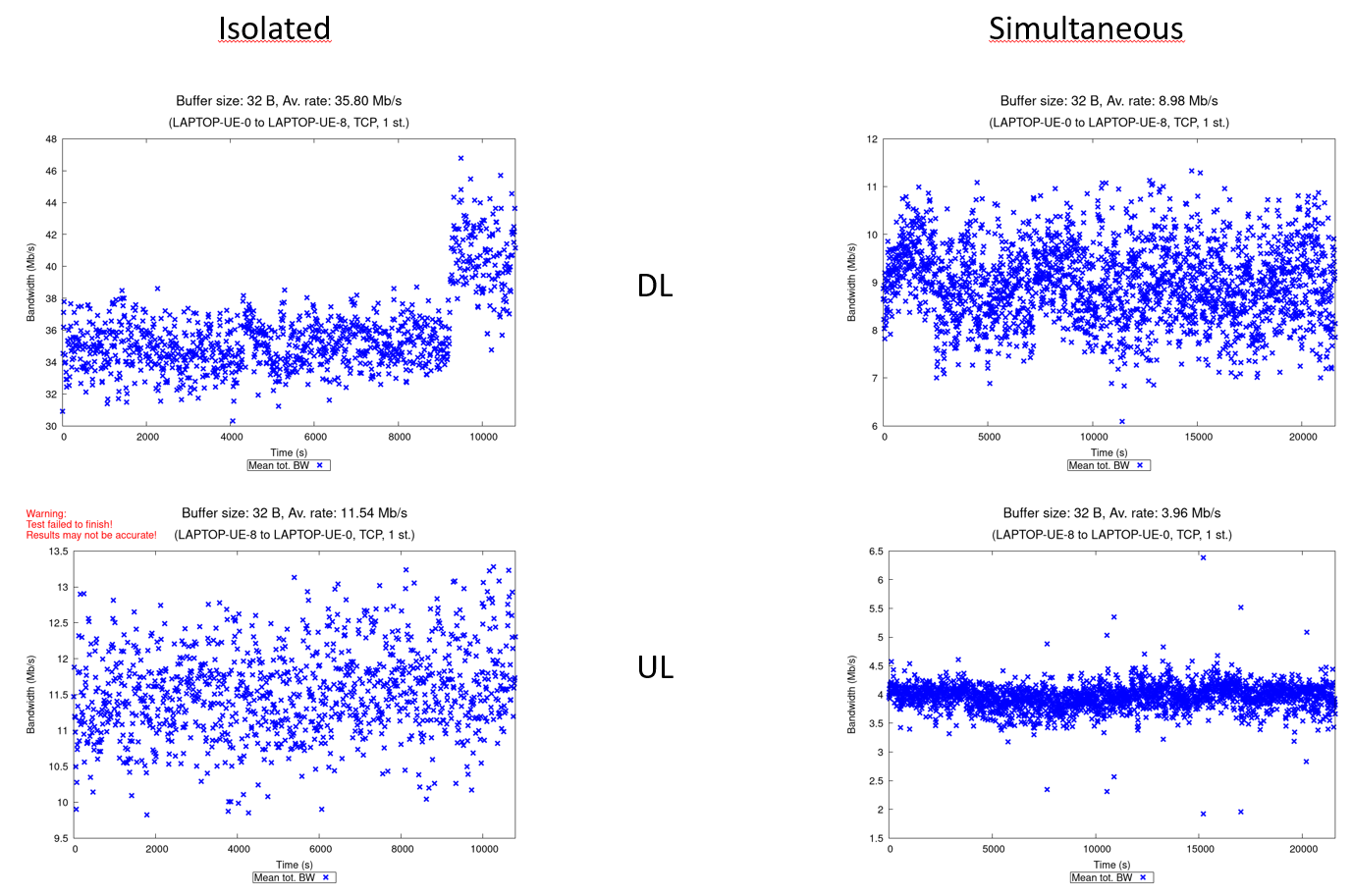}
    \caption{Performance evaluation carried out in the embedded system connected to the \ac{CPE} installed on the solar panel.}
    \label{fig:throughput_04}
\end{figure}

\begin{figure}[hbt!]
    \centering
    \includegraphics[height=7.5cm]{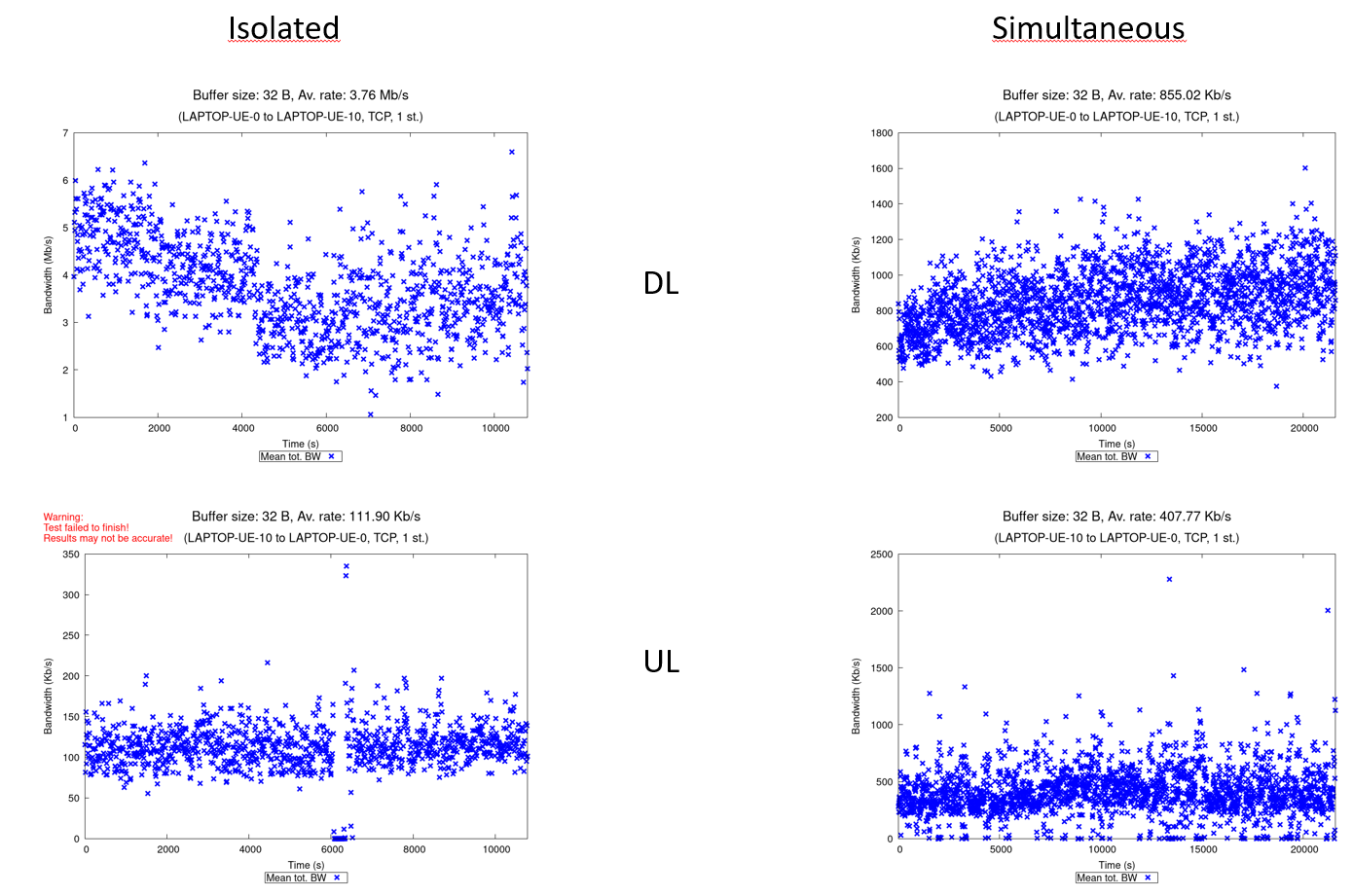}
    \caption{Performance evaluation carried out in the embedded system connected to the \ac{CPE} installed at Inverter 03.}
    \label{fig:throughput_05}
\end{figure}

In Figures \ref{fig:throughput_01}, \ref{fig:throughput_02}, \ref{fig:throughput_03}, \ref{fig:throughput_04}, and \ref{fig:throughput_05}, the isolated and simultaneous analyses of the performance are described for both uplink and downlink. 

\subsection{ICMP Performance Evaluation} 
\begin{figure}[hbt!]
    \centering
    \includegraphics[height=7.5cm]{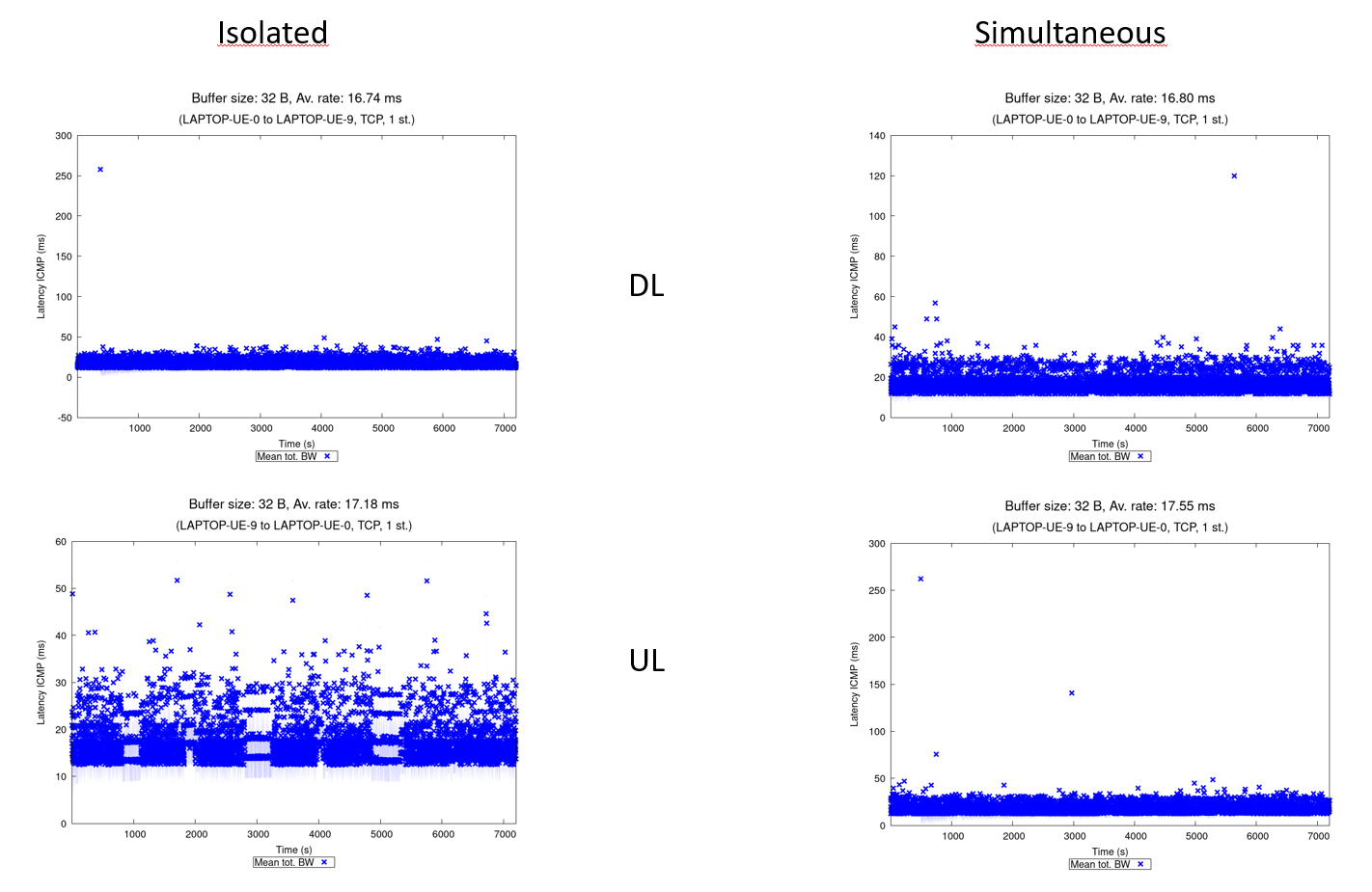}
    \caption{\ac{ICMP} analysis between the embedded system connected to the \ac{CPE} installed at the rectifier and PC0 connected to the core network.}
    \label{fig:icmp_01}
\end{figure}

\begin{figure}[hbt!]
    \centering
    \includegraphics[height=7.5cm]{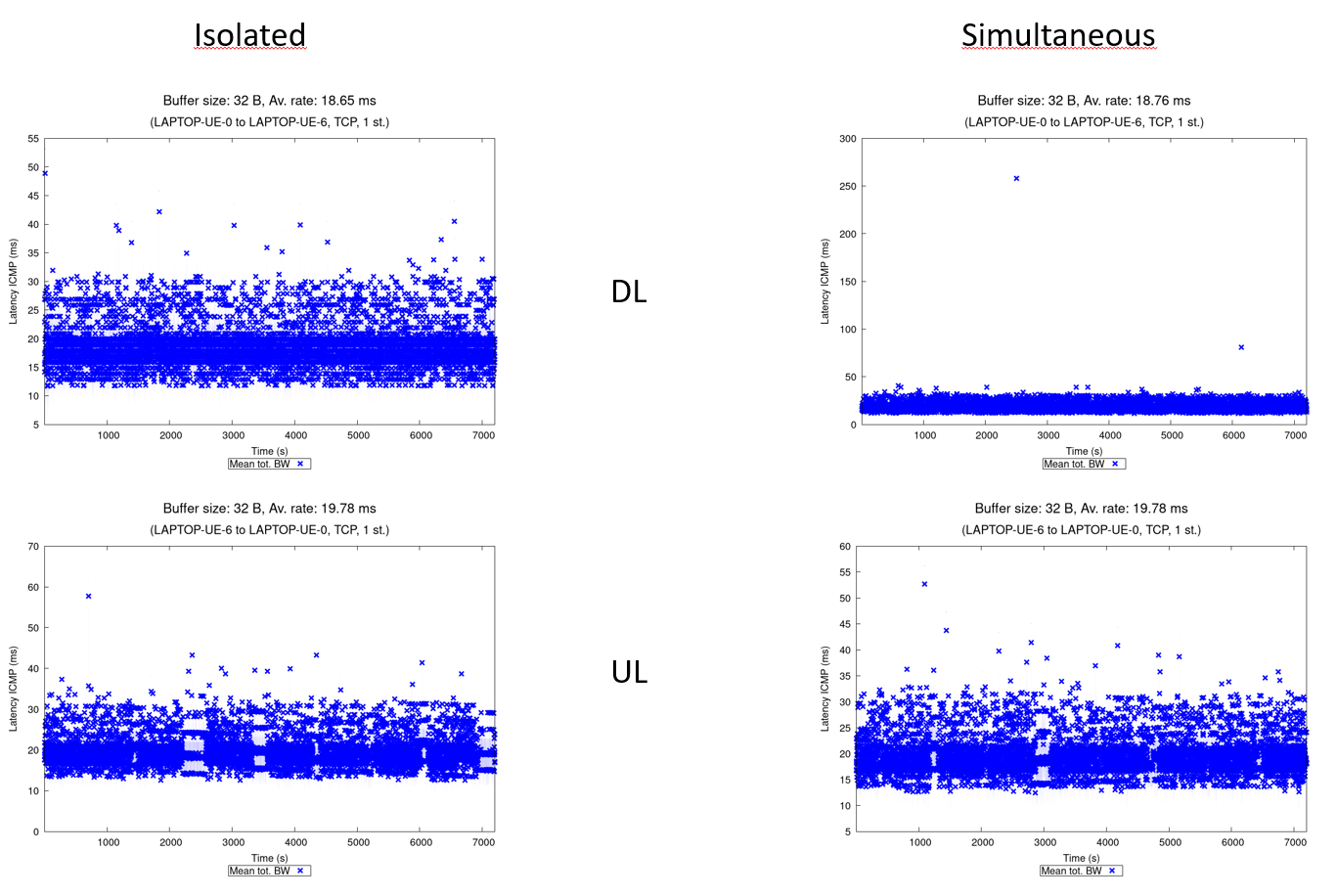}
    \caption{\ac{ICMP} analysis between the embedded system connected to the \ac{CPE} installed at Inverter 01 and PC0 connected to the core network.}
    \label{fig:icmp_02}
\end{figure}

\begin{figure}[hbt!]
    \centering
    \includegraphics[height=7.5cm]{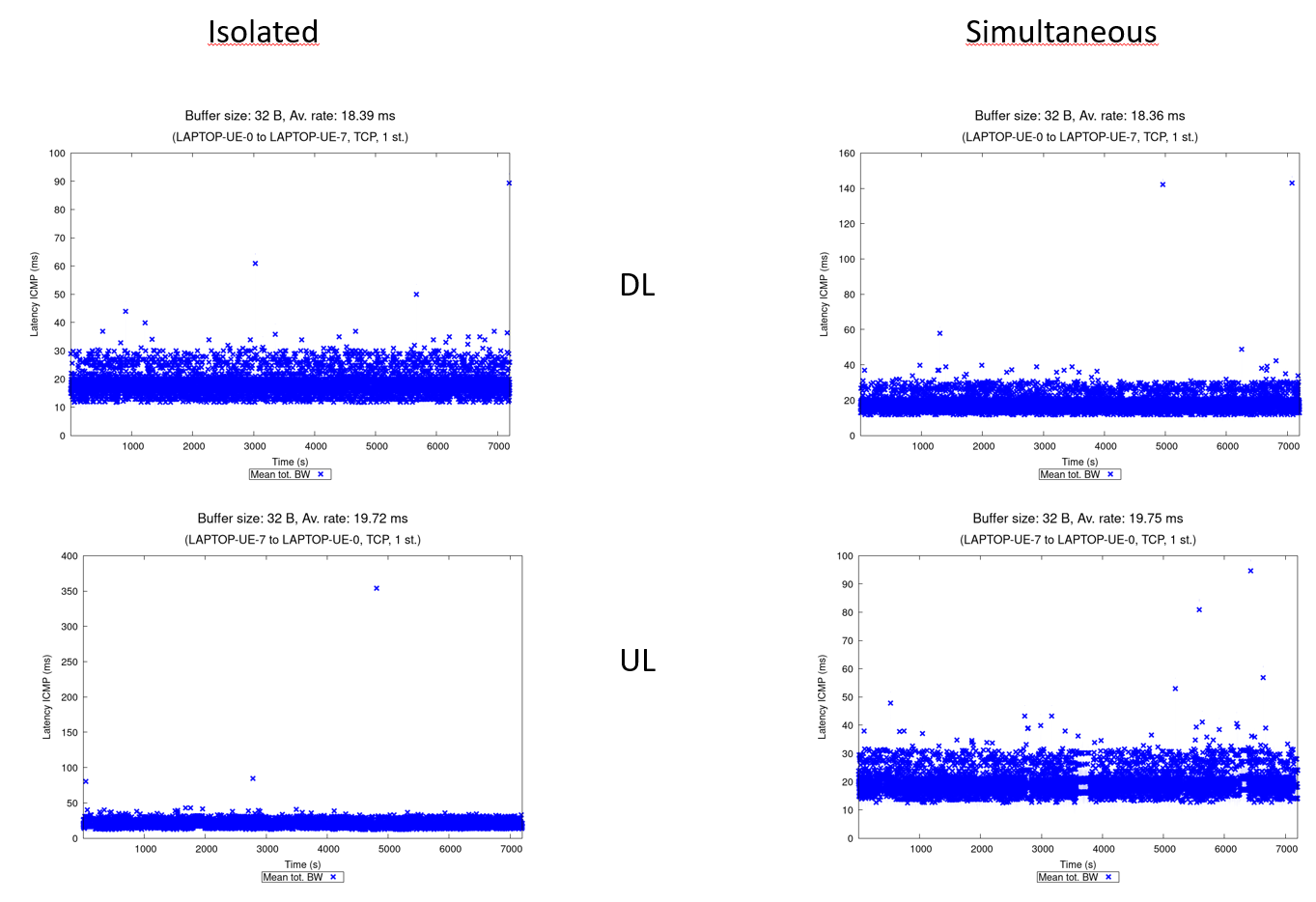}
    \caption{\ac{ICMP} analysis between the embedded system connected to the \ac{CPE} installed at Inverter 02 and PC0 connected to the core network.}
    \label{fig:icmp_03}
\end{figure}

\begin{figure}[hbt!]
    \centering
    \includegraphics[height=7.5cm]{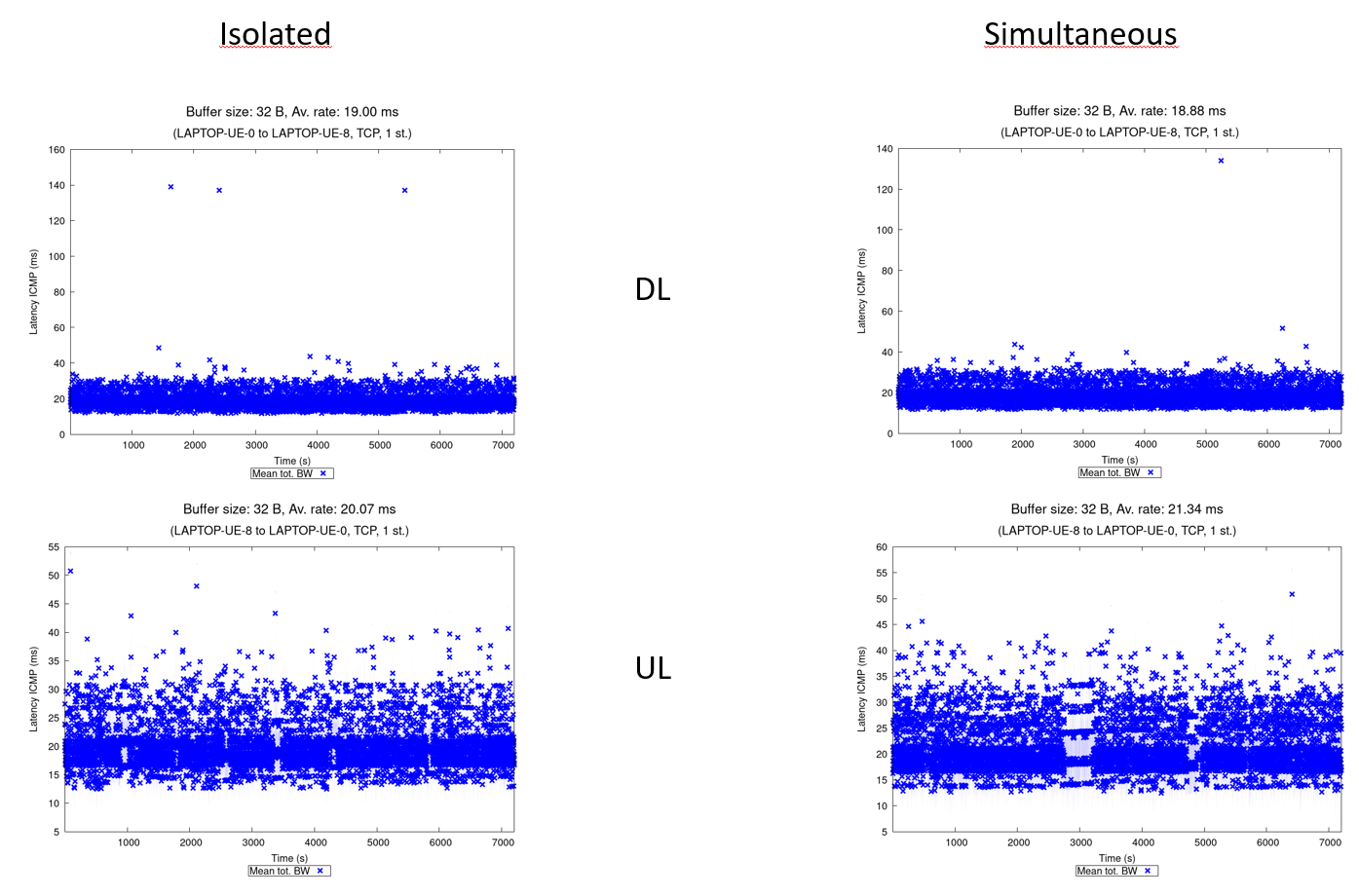}
    \caption{\ac{ICMP} analysis between the embedded system connected to the \ac{CPE} installed on the solar panel and PC0 connected to the core network.}
    \label{fig:icmp_04}
\end{figure}

\begin{figure}[hbt!]
    \centering
    \includegraphics[height=7.5cm]{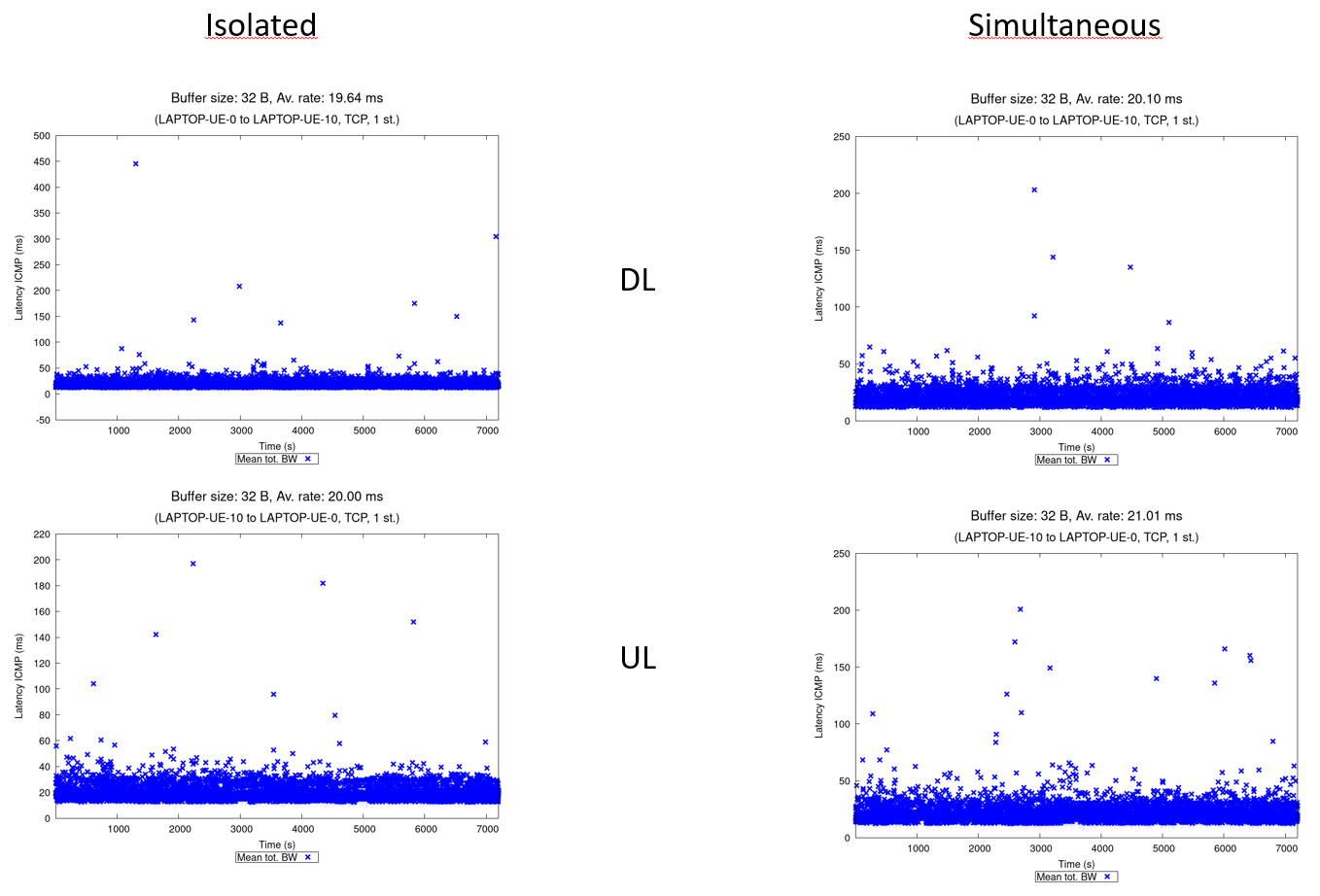}
    \caption{\ac{ICMP} analysis between the embedded system connected to the \ac{CPE} installed at Inverter 03 and PC0 connected to the core network.}
    \label{fig:icmp_05}
\end{figure}

Here, the isolated and simultaneous analysis was done based on the \ac{ICMP} protocol that measures the round-trip time transmission of the data between the transmitter and the receiver. 
Similar to the previous case, a specialized script was used to automate this measurement. The results for each of the five points are given in Figure \ref{fig:icmp_01}, \ref{fig:icmp_02}, \ref{fig:icmp_03}, \ref{fig:icmp_04}, and Figure \ref{fig:icmp_05}.

\subsection{Performance Evaluation of the Application}
\begin{figure}[hbt!]
    \centering
    \includegraphics[height=5.5cm]{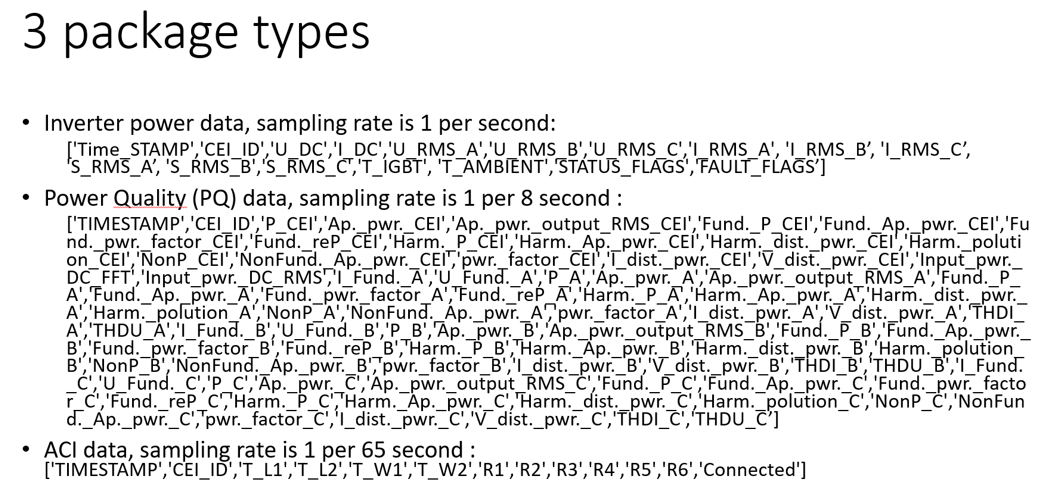}
    \caption{Description of the messages transmitted from each \ac{IED} following the database structure.}
    \label{fig:app_02}
\end{figure}

As described above, the application is based on the telemetry transmission of data gathered in each \ac{IED} installed in each microgrid element. 
These devices transmit over \acs{TCP} and \acs{UDP} \footnote{TCP and \acs{UDP} are the transport layer elements used by the standard IEC-61850.}-relevant information using different sample times between 1 sample per second and 1 sample every 65 s.
The messages transmitted to the database are described in Figure \ref{fig:app_02}.


\begin{figure}[hbt!]
    \centering
    \includegraphics[height=5cm]{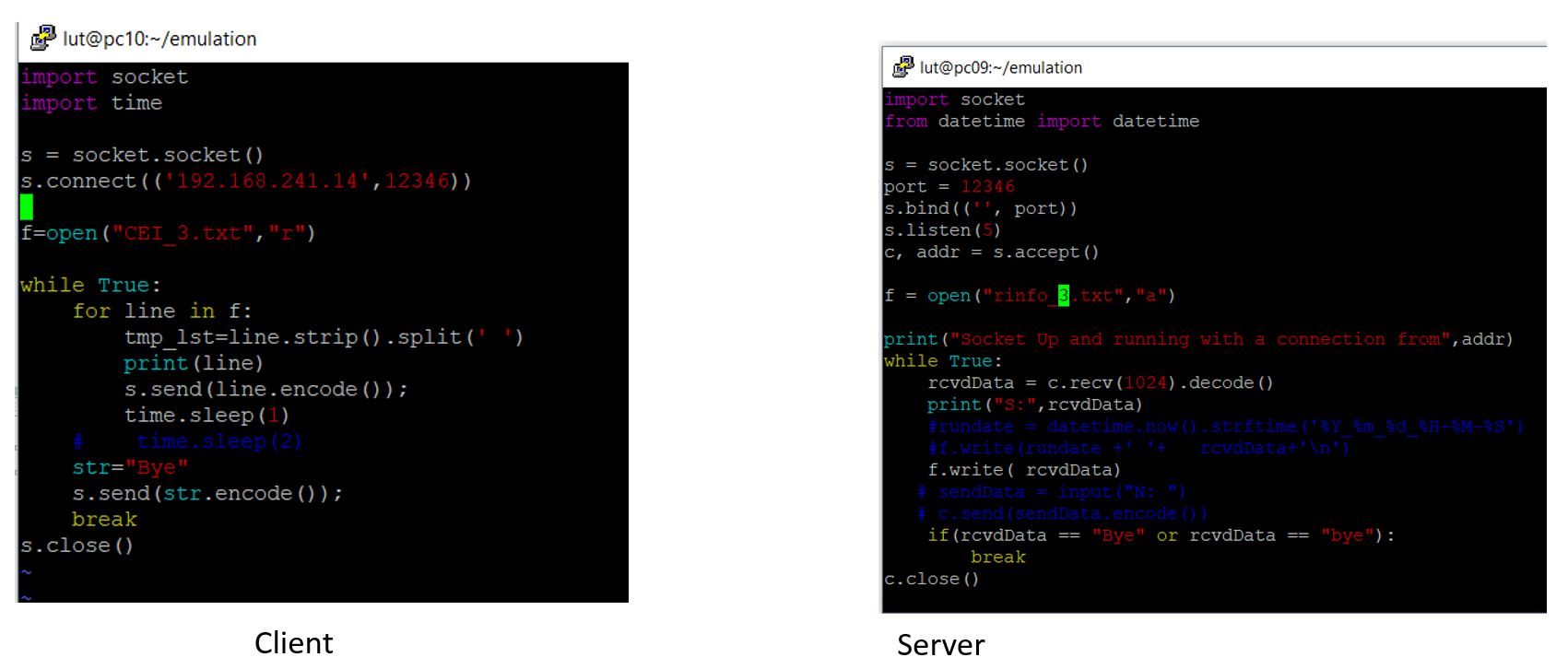}
    \caption{Part of the script used in the client-server application used in the test bed.}
    \label{fig:app_03}
\end{figure}

The application transmission sequence was emulated with a \ac{TCP} client-server application. These scripts enable to configure customized sample times for each scenario. A similar script was used to evaluate \ac{UDP}.
The scripts are illustrated in Figure \ref{fig:app_03}.

In this specific analysis there are different performance outputs. 
The first one is related to a simultaneous analysis of the message transmission from each inverter to the database using only TCP.
As described in Table \ref{tab:app_04}, the network obtained 0\% of data lost in all messages.

\begin{table}[hbt!]
     \caption{Message lost during simultaneous transmission of different messages, which are transmitted using different sampling times (between 1 and 65 s for each message transmission).}
    \centering
    \begin{tabularx}{\textwidth}{X|X|X|X}
    \hline
         &TCP - D1 & TCP - D2 & TCP - D3  \\
         \hline
       Inverter 01  &  0\% &  0\% &  0\% \\
       \hline
       Inverter 02  &  0\% &  0\% &  0\% \\
       \hline
       Inverter 03  &  0\% &  0\% &  0\% \\
       \hline
    \end{tabularx}
   
    \label{tab:app_04}
\end{table}

Even though the application uses only \ac{TCP}, based on the flexibility of the Python client-server script, the evaluation was extended to use \ac{UDP}.
This evaluation only focused on the message that is transmitted every second. 
Thus, in Table \ref{tab:app_05} it is possible to visualize that TCP is always getting a data message loss of 0\%. 
However, in the scenario with \ac{UDP}, some marginal error rate is obtained.

\begin{table}[hbt!]
    \caption{Message lost during simultaneous transmission of different messages, which are transmitted using different sampling times (between 1 and 65 s for each message transmission).}
    \centering
    \begin{tabularx}{\textwidth}{X|X|X}
    \hline
         &TCP (Simultaneous) - Error Rate (\%) & UDP (Simultaneous) - Error Rate (\%)  \\
         \hline
       Inverter 01  &  0\% &  0.2178\% \\
       \hline
       Inverter 02  &  0\% &  0.21669\%  \\
       \hline
       Inverter 03  &  0\% &  0.2777\%  \\
       \hline
    \end{tabularx}
    
    \label{tab:app_05}
\end{table}

\subsection{Other Results}
\begin{figure}[hbt!]
    \centering
    \includegraphics[height=7.5cm]{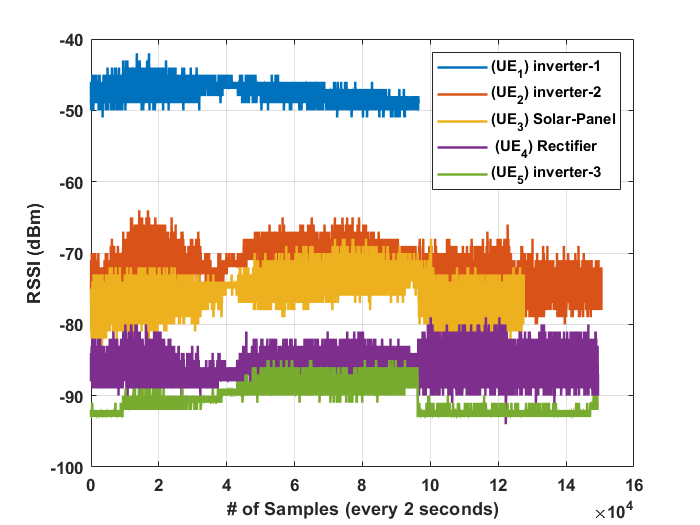}
    \caption{RSSI.}
    \label{fig:rf_rssi}
\end{figure}

\begin{figure}[hbt!]
    \centering
    \includegraphics[height=7.5cm]{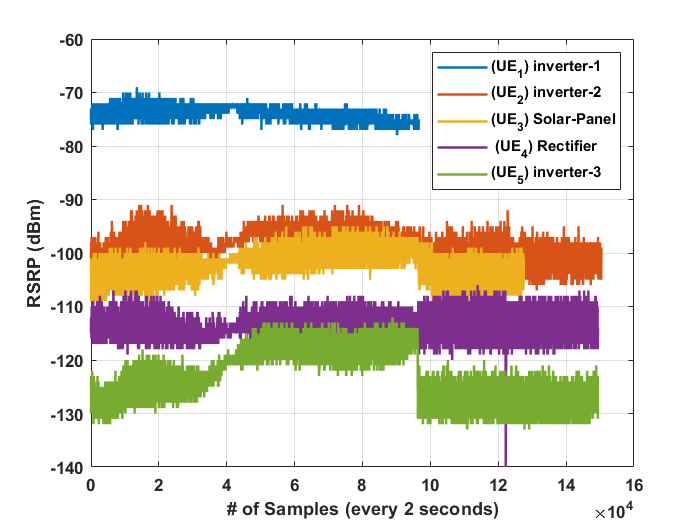}
    \caption{RSRP.}
    \label{fig:rf_rsrp}
\end{figure}

\begin{figure}[hbt!]
    \centering
    \includegraphics[height=7.5cm]{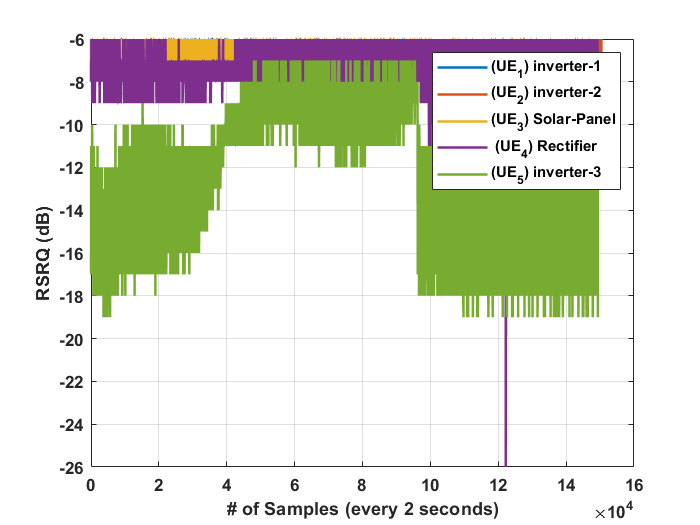}
    \caption{RSRQ.}
    \label{fig:rf_rsrq}
\end{figure}

\begin{figure}[hbt!]
    \centering
    \includegraphics[height=7.5cm]{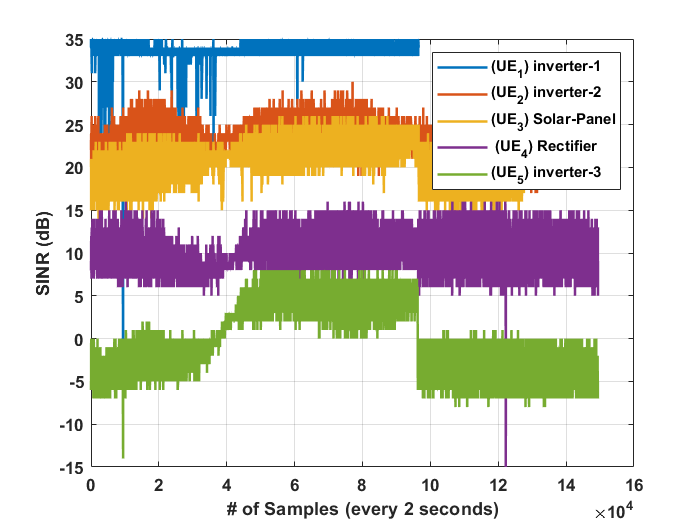}
    \caption{SINR.}
    \label{fig:rf_sinr}
\end{figure}

One interesting output of the field measurements is the time-varying evaluation of specific radio metrics, such as \ac{RSRP}, \ac{RSRQ}, \ac{SINR}, and \ac{RSSI}. 
These metrics are shown in Figure \ref{fig:rf_rssi}, \ref{fig:rf_rsrp}, \ref{fig:rf_rsrq}, and \ref{fig:rf_sinr}.

\newpage
\section{Discussion}

The main objective of this test bed was to evaluate the performance of a \ac{5G} private cellular network.
The idea was to examine the readiness of the \ac{TCP} and \ac{UDP} stack to transport specific telemetry information, which was gathered from different points in a real microgrid. 
\ac{TCP} and \ac{UDP} were considered, because they are the transport layer protocols defined in the standard IEC-51850.
Based on these assumptions, different  performance analyses were conducted to evaluate the throughput, latency, and data message integrity in the application. 

The throughput was evaluated in both scenarios; the first approach considered a simultaneous transmission, and the other one an isolated transmission for each node.
Here, the throughput was between 111.9 Kbps (node in the edge; Inverter 03) and the most robust node that achieved 85 Mbps, considering both scenarios.
In general, the simultaneous scenario reduced the throughput in each node. 
However, in the case of the edge node (Inverter 03), the uplink outperformed the simultaneous setup when compared with the isolated scenario.
This performance is probably related to the diversity and self power control of the \ac{RAN} when multiple nodes are being used.

In the case of latency, the \ac{ICMP} measurement indicates that in all cases (simultaneous and isolated), including nodes in the edge of the network gave an average latency of 20 ms, which is the maximum latency based on the numerology available in LTE-Advanced specifications.

In the case of the application analysis, the message integrity evaluation indicated that when using TCP, the wireless network achieves 0\% of lost. However, this performance was slightly degraded when \acs{UDP} was used as the transport protocol.

\newpage
\bibliographystyle{plainnat}
\bibliography{bibliography/references.bib}
\end{document}